\documentclass[a4paper,11pt]{article}
\usepackage{jheppub}
\usepackage{mathrsfs}
\usepackage{amsfonts}
\usepackage{setspace}
\usepackage{cellspace}
\usepackage{amsmath,bm}
\usepackage[colorlinks=true,linkcolor=blue]{hyperref}
\usepackage{xcolor}
\usepackage{epsfig}
\usepackage{slashed}
\usepackage{caption}
\usepackage{hhline,multirow,tabularx}  % for nicer tables
\usepackage{dcolumn}    % align table columns on decimal point
\usepackage{url}        % for URL addresses
\usepackage{braket}

\title{Spin-operator form factors of the critical Ising chain and their finite volume scaling limits}
\author[a]{Yizhuang Liu}

\affiliation[a]{Institute of Theoretical Physics,
Jagiellonian University, 30-348 Kraków, Poland}

\emailAdd{yizhuang.liu@uj.edu.pl}

\abstract {In this work, we provide a self-contained derivation of the spin-operator matrix elements in the fermionic basis, for the critical periodic Ising chain at a generic system length $N\in 2Z_{\ge 2}$. The approach relies on the near-Cauchy property of certain matrices formed by the Toeplitz symbols in the critical model, and leads to a few square-root products for the leg functions. The square root products allow simple integral representations, that further reduce to the Binet's second integral and its generalization by Hermite,  in the {\it finite volume scaling limit}. This leads to product formulas for the spin operator matrix elements in the scaling limit, providing explicit expressions for the spin-operator form factors of the Ising CFT in the fermionic basis, that were computed iteratively in~\cite{Yurov:1991my}. They are all rational numbers up to $\sqrt{2}$. We also determine the normalization factor of the spin-operator and show explicitly how the coefficient $G(\frac{1}{2})G(\frac{3}{2})$ appear through a ground state overlap. Moreover, by expanding the spin-spin two point correlator in the fermionic basis, we observed a Fredholm determinant identity
\begin{align}
\det (1+{\cal  K}_{ij}(w))_{1\le i,j\le \infty}=(1-w)^{-\frac{1}{8}},  \  \ {\cal K}_{ij}(w)=\frac{1}{\pi}\frac{i}{i+j}\frac{\Gamma(i+\frac{1}{2})\Gamma(j+\frac{1}{2})}{\Gamma(i+1)\Gamma(j+1)} w^j  \ ,  \label{eq:abs}
\end{align}
which allows to show the convergence of the rescaled two-point correlator to the CFT version on a cylinder.}
\date{\today}

\begin{document}
\maketitle
\flushbottom
\section{The critical Ising chain, square-root products and main results}
The finite volume spin-operator form factors of the massive fermion Ising-QFT~\cite{Fonseca:2001dc} were well known, but what about their massless limits? Will the results become meaningless, or will they reduce to the spin-operator matrix elements of the Ising CFT in the fermionic basis, that were computed only iteratively in~\cite{Yurov:1991my}? To the author's knowledge, it seems that many references who used the massive form factors in~\cite{Fonseca:2001dc}, including this paper itself, forgot to mention their convergence to a nontrivial finite volume massless scaling limit. Instead, the integrals in the leg functions were usually written in certain forms, creating impressions of  their ill-posedness or triviality at the zero fermion mass\footnote{For example, in Ref.~\cite{Gabai:2019ryw}, the leg function was written with an overall $m^2$. Still, it is easier to spot a  non-trivial massless limit from the representation in~\cite{Gabai:2019ryw} than the rapidity space integral in~\cite{Fonseca:2001dc}.}. 

Similarly, in the literature, there are plenty discussions and derivations~\cite{Bugrij:2001nf,Lisovy:2001gc,Bugrij:2003xhk,Bugrij:2004sac,Gehlen_2008,Iorgov:2010yv,Iorgov:2010ie} for the finite volume form factors in the Ising model and the Ising chain. These works are very non-trivial, but most of them were assuming $T>T_c$ or $T<T_c$, and forgot to mention what happened at the critical point. For example, in the published version of~\cite{Gehlen_2008}, the authors said explicitly after the Eq.~(78) that the results are valid both when $g>1$ and $0\le g< 1$, but just omitted the critical point $g=1$. Very often, the formulas were written with a refactor such as $\xi=(g^2-1)^{\frac{1}{4}}$ in front, a factor that vanishes at the critical point $g=1$, like the expressions given to the leg functions in the massive theory, despite that in the proper way of taking the $g\rightarrow 1^+$ limit, the results are actually finite and as we will show, lead exactly to the form factors at the critical point. 

Meanwhile, in the literature, there were also not too many discussions on the spin-operator form factors in the Ising-CFT, despite that they are clearly universal and unique quantities: the massless fermion basis of the Ising CFT is perhaps among the few integrable massless scattering basis~\cite{Bazhanov:1994ft,Bazhanov:1996aq} of 2D CFTs, that allow explicit expressions for local operator's matrix elements between them. On the other hand, in comparison to the Virasoro basis, it could be the massless scattering basis, that respect better the scale-full deformations of CFTs either to the integrable lattice models in the UV~\cite{Boos:2010qii}, or to the massive QFTs in the IR. It is also possible, that for certain non-inclusive high-energy limits that are more sensitive to the structures of the Hilbert space, one needs not just the standard CFT perturbation theory for point-like objects, but also the correct massless basis, to write out the correct high-energy asymptotics. These questions deserve more investigations. Furthermore,  to the author's taste, CFTs are beautiful not just for themselves, but also for the explicit asymptotics telling how they are approached. It is always interesting to establish such asymptotics from the first principle for various quantities, whenever possible. 

Given all above, the author believes that it is not unreasonable to provide a self-contained derivation for the spin-operator form factors in the 2D periodic Ising chain, directly at the critical point. Below we will provide such a derivation, and show how in the finite volume scaling limit, the results converge to the Ising CFT versions. We believe that many of the products, integral representations, large $N$ asymptotics and the matrix elements in this work are novel, beautiful and simple. We also believe that the explicit formulas provided by this work could be of practical usages (such as the TFCSA applications~\cite{Yurov:1991my}).

\subsection{The square root function and roots of $\pm1$}
 We first clarify the notations used in the work. Throughout the work, unless otherwise stated, we use the $(0,2\pi)$ branch of the square-root function
\begin{align}
\sqrt{{\cal Z}}={\cal Z}^{\frac{1}{2}}=|{\cal Z}|^{\frac{1}{2}}e^{\frac{i}{2}{\rm Arg}({\cal Z})} \ , \  0<{\rm Arg}({\cal Z})<\pi \ , \\
\sqrt{{\cal Z}}=-{\cal Z}^{\frac{1}{2}}=-|{\cal Z}|^{\frac{1}{2}}e^{\frac{i}{2}{\rm Arg}({\cal Z})} \ , \  -\pi<{\rm Arg}({\cal Z})<0 \ . 
\end{align}
As such, $\sqrt{{\cal Z}}$ is an analytic function with a branch cut along the positive real axis. This choice is natural in the context of the critical Ising chain. Without otherwise mentioning, for positive numbers $p\in R_{\ge0}$, we use the upper branch with $\sqrt{p}>0$. We choose the system size $N \in 2Z_{\ge 2}$ to be a positive even integer. Given $N$, the sets of the roots of $\pm 1$ are denoted as $S_\pm$
\begin{align}
S_+=\{\beta;  \ \beta^N=1\} \ , \ S_-=\{z; \  z^N=-1\} \ . 
\end{align}
They are the exponentials of the periodic and anti-periodic momenta. {\it Throughout this paper, the small letter $\beta$ in formulas will be reserved for $S_+$, and the small letter $z$ reserved for $S_-$.} Since $N$ is even, $-1 \in S_+$, and  points in $S_+$ except $1,-1$ can be grouped to $\frac{N-2}{2}$ pairs symmetric with respect to the real axis.  Similarly, all points in $S_-$ are grouped to $\frac{N}{2}$ such pairs. Using these properties, one has first of two of the product formulas to be used latter
\begin{align}
\prod_{\beta \in S_+} \sqrt{\beta}=i (-1)^{\frac{N-2}{2}} \ , \prod_{z\in S_-} \sqrt{z}= (-1)^{\frac{N}{2}} \ , \label{eq:productsimple}
\end{align}
where we will use the upper branch  $\sqrt{1}=1$ at $\beta=1$ without otherwise mentioning. We also need the following product formulas  
\begin{align}
\prod_{\beta \in S_+;  \ \beta \ne 1} \frac{\sqrt{\beta}+1}{\sqrt{\beta}-1} =\frac{i+1}{i-1}(-1)^{\frac{N-2}{2}} \ ,  \ \prod_{z\in S_-} \frac{\sqrt{z}-1}{\sqrt{z}+1}=(-1)^{\frac{N}{2}} \ . \label{eq:prod}
\end{align}
This is because in the $(0,2\pi)$ branch, for $z\ne 1$ one has $\sqrt{\frac{1}{{\cal Z}}}=-\frac{1}{\sqrt{{\cal Z}}} $, 
and for $e^{i\alpha} \ne \pm 1$,  one has 
\begin{align}
\frac{e^{i\alpha}-1}{e^{i\alpha}+1}\times \frac{-e^{-i\alpha}-1}{-e^{-i\alpha}+1}=-1 \ . 
\end{align}
Latter we will use these formula to simplify the leg functions at $\beta=1$. 

\subsection{The critical Ising chain}

After introducing the sets $S_\pm$, we set up the conventions for the critical periodic Ising chain. We choose the $-, +$ convention of the Hamiltonian 
\begin{align}
H=-\frac{1}{2}\sum_{i=0}^{N-1}\sigma^x_i\sigma^x_{i+1}+\frac{1}{2}\sum_{i=0}^{N-1}\sigma^z_i \ ,  \  \sigma^x_{N}=\sigma^x_0 \ .  \label{eq:Hamil}
\end{align}
Here, $\sigma_i^x,\sigma_i^y,\sigma_i^y$ are the standard $2\times 2$ Pauli matrices at the site $i$. These spin operators commute at different sites. It is possible to choose the $\sigma^x$ and $\sigma^z$ to be purely real, as such $H$ can also be purely real.  We define the standard Dirac matrices as~\cite{Jordan:1928wi,Kaufman:1949ks} 
\begin{align}
&\Gamma_{2i}=\sigma_i^x\prod_{0\le j<i}  (-\sigma^z_i) \ , \Gamma_{2i+1}=-\sigma_i^y\prod_{0\le j<i} (-\sigma^z_i) \ , \   0\le i \le N-1 \ , \label{eq:fermionrel}\\ 
&\{\Gamma_i,\Gamma_j\}=2\delta_{ij} \ , \ 0\le i,j \le 2N-1 \ . 
\end{align}
In terms of them, one has the relations
\begin{align}
& \sigma_i^z =i\Gamma_{2i}\Gamma_{2i+1} \ , \   0\le \ i\le N-1 \ ;   \ \sigma^x_i\sigma^x_{i+1}=-i\Gamma_{2i+1}\Gamma_{2i+2}\ , \  0\le  i\le N-2 \ ,  \\
& \sigma^x_{N-1}\sigma^x_0=i\Gamma_{2N-1}\Gamma_0\prod_{i=0}^{N-1}(-i)\Gamma_{2i}\Gamma_{2i+1}\ .
\end{align}
As such, in terms of the Dirac matrices, the Hamiltonian reads
\begin{align}
H=\frac{i}{2}\sum_{i=0}^{2N-2}\Gamma_{i}\Gamma_{i+1}-\frac{i}{2}\Gamma_{2N-1}\Gamma_0 Z\ ,  \  Z=\prod_{i=0}^{N-1}(-i)\Gamma_{2i}\Gamma_{2i+1} \ . \label{eq:fermH}
\end{align}
The operator $Z$ is just the standard $Z_2$ charge and commutes with $H$ and other even combinations of fermion operators. As such, in the charge odd and even sectors, one has two different quadratic Hamiltonians, but in terms of the same set of fermion operators.  In each of the sectors, one can diagonalize the Hamiltonian by finding the proper, orthogonal transformations that diagonalize the skew-symmetric matrices appearing in $H$.

To further set up our conventions for the eigensystem,  especially, to convince the readers that our sign choices at the $\beta=1$ point are correct, here we review how the diagonalization process works in the $Z=-1$ (R) sector. The $Z=1$ (NS) sector can be handled similarly and is free from the $\beta=1$ issues. We will mainly follow the spirit of~\cite{Kaufman:1949ks}, but the computations are simpler in the Ising chain than the Ising model. For $Z=-1$, one has the following rotation matrix 
\begin{align}
&{ \cal E}_{ij}=\delta_{j,i+1}-\delta_{i,j+1}  \ ,  \ 0 \  \le i, j \le 2N-1 \  \ ; \  {\cal E}_{2N-1,0}=-{\cal E }_{0,2N-1}=1 \ , \\
&H(1-Z)=\frac{i}{4}\sum_{i,j}\Gamma_i\Gamma_j {\cal E}_{ij} (1-Z)\ . 
\end{align}
To diagonalize the Hamiltonian, one needs to find the orthogonal transformation $\Omega$ such that the $\Omega^T{\cal E}\Omega$ is block-diagonalized. It is sufficient to find the normalized eigenvectors $\lambda(j)$ of ${\cal E}$ with the eigenvalues $i\epsilon(j)$ where $\epsilon(j)>0$, and fill the $2j$-th and the $2j+1$-th columns of $\Omega$ by $\sqrt{2}{\rm Re}(\lambda(j))$ and $\sqrt{2}{\rm Im}(\lambda(j))$. To find them, we write the eigenvalue equation as
\begin{align}
&i\epsilon \lambda_i=\lambda_{i+1}-\lambda_{i-1} \ , 1\le i\le 2N-2  \ , \\
&i\epsilon \lambda_0=\lambda_1-\lambda_{2N-1} \ , i\epsilon\lambda_{2N-1}=\lambda_0-\lambda_{2N-2} \ . 
\end{align}
To be compatible with the results away from the critical point, we treat the even and odd sites differently and write the ansartz as 
\begin{align}
&\lambda_{2k}(\beta)=\beta^k u(\beta) \ ,  \  \lambda_{2k+1}(\beta)=\beta^k\nu(\beta) \ , \  \beta^{N}=+1 \ ,  \label{eq:vectlambda} \\ 
&i\epsilon(\beta) u(\beta)=\nu(\beta)(1-\bar \beta) \ ,  \ i\epsilon(\beta)\nu(\beta)=u(\beta)(\beta-1) \ . 
\end{align}
For $\beta \ne 1$, the above can be solved as 
\begin{align}
&\epsilon(\beta)=-i\left(\sqrt{\beta}-\frac{1}{\sqrt{\beta}}\right)>0 \ ; \  u(\beta)=i \ ,  \ \nu(\beta)=\frac{\beta-1}{\epsilon(\beta)}=i\sqrt{\beta} \  . \label{eq:choiceunu}
\end{align}
Notice that the $\sqrt{\beta}$ here is in the $(0,2\pi)$ branch. For $\beta=1$, although $\epsilon(1)=0$, there are more freedoms for the eigenvectors. Here we  introduce the function
\begin{align}
\varphi(\beta)=\sqrt{\beta} , \  \beta \ne 1 ;  \ \varphi(1)=i \ , 
\end{align}
and make the following choice to be justified latter
\begin{align}
u(1)=i \ ,  \  \nu(1)=i\varphi(1)=-1 \ .
\end{align}
We then introduce the $2N \times 2N$ rotation matrix $\Omega$. The rows of $\Omega$ are labeled by the indices of the gamma matrices ordered from $0$ to $2N-1$, while the columns are labeled by the $\beta \in S_+$. For each $\beta$, there are two columns, one denoted by $\beta$, and another denoted by $\beta+\frac{1}{2}$. We always place the $\beta+\frac{1}{2}$ column right next to the $\beta$ column. The elements of $\Omega$ read
\begin{align}
&\Omega_{2k,\beta}=\frac{1}{\sqrt{N}}{\rm Re}(i\beta^k), \   \Omega_{2k,\beta+\frac{1}{2}}=\frac{1}{\sqrt{N}}{\rm Im}(i\beta^k) \ , \label{eq:rotat1}\\
&\Omega_{2k+1,\beta}=\frac{1}{\sqrt{N}}{\rm Re}(i \varphi(\beta)\beta^k), \   \Omega_{2k+1,\beta+\frac{1}{2}}=\frac{1}{\sqrt{N}}{\rm Im}(i\varphi(\beta)\beta^k) \  .\label{eq:rotat2}
\end{align}
The factor $\frac{1}{\sqrt{N}}$ is because the norm of the unnormalized vector $\lambda(\beta)$ in Eq.~(\ref{eq:vectlambda}) is $\sqrt{2N}$. From the rotation matrix, one can form the rotated Dirac matrices
\begin{align}
&\gamma_\beta=\sum_{k=0}^{N-1}\bigg(\Gamma_{2k}\Omega_{2k,\beta}+\Gamma_{2k+1}\Omega_{2k+1,\beta} \bigg)\ , \label{eq:gammab} \\   
&\gamma_{\beta+\frac{1}{2}}=\sum_{k=0}^{N-1}\bigg(\Gamma_{2k}\Omega_{2k,\beta+\frac{1}{2}}+\Gamma_{2k+1}\Omega_{2k+1,\beta+\frac{1}{2}}\bigg) \ ,\label{eq:gammbhf}
\end{align}
that block diagonalize the Hamiltonian under the $1-Z$ projection 
\begin{align}
H\frac{1-Z}{2}=\frac{i}{2}\sum_{\beta\in S_+}\epsilon(\beta) \gamma_\beta\gamma_{\beta+\frac{1}{2}} \frac{1-Z}{2}\ . 
\end{align}
At this point, one only knows that $\Omega^T\Omega=1$, but to further determine the spectrum, one needs to know the sign of $\det \Omega$ in order to express the charge operator $Z$ in terms of the $
\gamma_{\beta}, \gamma_{\beta+\frac{1}{2}}$s.

Instead of computing the determinant by brute force, we deform $\Omega$ by replacing the $\varphi(\beta)$ with
\begin{align}
\varphi(\beta,h)=i\left(\frac{1-h\beta}{1-h\bar \beta}\right)^{\frac{1}{2}} \ ,  \  0 \le h<1 \ , \label{eq:defphig}
\end{align}
where the square-root function is defined with the principal branch. The deformed matrices, called $\Omega(h)$,  are still orthogonal and are for the Ising chain in the disordered region, whose Hamiltonian contains an extra $h$ in front of the $\sigma^x_i\sigma^x_{i+1}$ terms. Furthermore, $\Omega(h)$ is a continuous function of $h$ when $0\le h<1$ and $\beta \in S_+$, and the $h\rightarrow 1^-$ limit exactly reduces to the rotation matrix $\Omega$ for the critical model. In particular, at $\beta=1$, one has 
\begin{align}
\lim_{h\rightarrow 1^-} \varphi(1,h)=i \ , 
\end{align}
justifying the choice made above. The continuity implies that $\det \Omega(1^-)=\det \Omega(0)$, since the determinant is a continuous function of $h$ which could take only two values $\pm1$. It is not hard to compute that $\det\Omega(0)=1$, since at $h=0$ the Hamiltonian is non-interacting, and $\Omega(0)$ is a $2N\times 2N$ rotation underlying an $N\times N$ unitary transformation. As such, the rotation matrix $\Omega=\Omega(1^-)$ is already proper. Notice if one choose $\varphi(1)=-i$, then the $\det \Omega$ would be $-1$, since this choice only changes the sign of $\gamma_0$.  Knowing the sign of $\det \Omega$, one can further introduce the creation and annihilation operators
\begin{align}
c_\beta^\dagger=\frac{1}{2}\left(\gamma_\beta-i\gamma_{\beta+\frac{1}{2}}\right) \ ,  \  c_\beta=\frac{1}{2}\left(\gamma_\beta+i\gamma_{\beta+\frac{1}{2}}\right) \ , \label{eq:cb}
\end{align}
in terms which the Hamiltonian under the $1-Z$ projection and the charge operator read
\begin{align}
&H\frac{1-Z}{2}=\sum_{\beta\in S_+}\epsilon(\beta)\left(c_\beta^\dagger c_\beta-\frac{1}{2}\right) \frac{1-Z}{2} \ , \\ 
& Z=\det \Omega \prod_{\beta\in S_+}(-i\gamma_\beta\gamma_{\beta+\frac{1}{2}})=\prod_{\beta\in S_+}(1-2c_\beta^\dagger c_\beta) \ . \label{eq:charge}
\end{align}
Since $\Omega$ is non-degenerate, one can always find a normalized state $|\Omega_R\rangle \ne 0$ annihilated by all the $c_\beta$s. This state must be charge even, due to the Eq.~(\ref{eq:charge}) for the $Z_2$ charge. Since the representation space is $2^N$ dimensional, the Fock space generated by acting the $c_\beta^\dagger$s on $|\Omega_R\rangle$ is the full space. Among them, only those with an odd number of $c_\beta^\dagger$s are energy eigenstates, due to the $1-Z$ projection.  The diagonalization process in the NS sector is similar. All the expressions in the R sector remain the same form, except $\beta\in S_+$ should be replaced by $z\in S_-$. There is no $z=1$ issue in NS and $\varphi(z)=\sqrt{z}$ for all $z\in S_-$. 

Combining the two sectors, the full Hamiltonian reads
\begin{align}
&H=\sum_{\beta\in S_+}\epsilon(\beta)\left(c_\beta^\dagger c_\beta-\frac{1}{2}\right) \frac{1-Z}{2} +\sum_{z\in S_-}\epsilon(z)\left(c_z^\dagger c_z-\frac{1}{2}\right) \frac{1+Z}{2} \ , \\
&Z=\prod_{\beta\in S_+}(1-2c_\beta^\dagger c_\beta)=\prod_{z\in S_-}(1-2c_z^\dagger c_z) \ . 
\end{align}
The state $|\Omega_{NS}\rangle$ annihilated by all the $c_z$s is the ground state in the charge even sector, sector, and the $c_1^{\dagger}|\Omega_R\rangle$ is the ground state in the charge odd sector. The $c_1^\dagger$ excitation will be called the zero mode. On the other hand, although $\epsilon(1)=0$, the energy difference $\Delta_E(N)$ between the R and NS ground states is non-zero and approaches $\frac{\pi}{4N}$ as $N \rightarrow \infty$. To show this, notice that 
\begin{align}
\Delta_E(N)=\frac{1}{2}\sum_{z\in S_-} \epsilon(z)-\frac{1}{2}\sum_{\beta \in S_+} \epsilon(\beta) \ . 
\end{align}
The sums are geometric and can be performed in elementary manner. It is also possible to use the weighting functions $\frac{1}{2\pi iz  }\frac{-N}{z^N+1}$ and $\frac{1}{2\pi i \beta }\frac{N}{\beta^N-1}$ to express the summations as contour integrals, and then deform the contours to the branch cuts along the positive real axis. At the end, one has
\begin{align}
\Delta_E(N)=-\frac{1}{2\pi}\int_0^1 dt \frac{1+t}{t\sqrt{t}}\ln \frac{1-t^N}{1+t^N}=\tan \frac{\pi}{4N}=\frac{\pi}{4N}+\frac{\pi^3}{192N^3}+{\cal O}\left(\frac{\pi^5}{N^5}\right) \ . \label{eq:deltae}
\end{align}
As such, the state $|\Omega_{NS}\rangle$ is the absolute ground state of the system, whose energy will be normalized to $0$. On the other hand,  the energy gap $\frac{\pi}{4N}$ is exactly due to the scaling dimension of the spin operator and allows the identification of the state $c_1^\dagger|\Omega_R\rangle$ as the primary state $|\hat \sigma\rangle$ of the spin operator in the CFT language.  On top of the ground states, a complete set of energy eigenstates of the system and their energies are given by
\begin{align}
&NS: \prod_{i=1}^{2k} c_{z_i}^{\dagger}|\Omega_{NS}\rangle,  \  E=\sum_{i=1}^{2k}-i\left(\sqrt{z_i}-\frac{1}{\sqrt{z_i}}\right)\ ; \label{eq:systemNS}\\
&R: \prod_{i=1}^{2k'+1}c_{\beta_i}^{\dagger}|\Omega_R\rangle \ , \ E=\Delta_E(N)+\sum_{i=1}^{2k'+1}-i\left(\sqrt{\beta_i}-\frac{1}{\sqrt{\beta_i}}\right). \label{eq:systemR} 
\end{align}
They are exactly the fermionic basis, in terms which we will compute the matrix elements of the spin operator. 

Equally important as the spectrum for the form factor computations are the mode decompositions of the $\Gamma_i$s. In terms of the following Toeplitz symbol\footnote{The choice is to be consistent with the symbol $C(z,h)=\left(\frac{1-h \bar z}{1-hz}\right)^{\frac{1}{2}}$ for the Toeplitz determinant of the $\sigma\sigma$ two point correlator in the ordered region.}
\begin{align}
&C(\bar \beta)=C^{\dagger}(\beta)=\frac{1}{C(\beta)}=-i\varphi(\beta) \ ,
\end{align}
the mode decompositions in terms of $c_\beta, c_\beta^\dagger$ follow by inverting the Eq.~(\ref{eq:gammab}), Eq.~(\ref{eq:gammbhf}) using $\Omega\Omega^T=1$
\begin{align}
\Gamma_{2i}=&\sum_{\beta\in S_+}\left(\Omega_{2i,\beta}\gamma_\beta+\Omega_{2i,\beta+\frac{1}{2}}\gamma_{\beta+\frac{1}{2}}\right)=\frac{i}{\sqrt{N}}\sum_{\beta \in S_{+}} (\beta^ic^{\dagger}_\beta-\beta^{-i}c_\beta) \ , \label{eq:mode1}\\
\Gamma_{2i+1}&=\sum_{\beta\in S_+}\left(\Omega_{2i+1,\beta}\gamma_\beta+\Omega_{2i+1,\beta+\frac{1}{2}}\gamma_{\beta+\frac{1}{2}}\right)\nonumber \\ 
&=-\frac{1}{\sqrt{N}}\sum_{\beta \in S_{+}} (\beta^{i}C(\bar \beta)c_\beta ^{\dagger}+\beta^{-i}C(\beta)c_\beta) \ ,  \label{eq:mode2}
\end{align}
Similarly, the same set of $\Gamma_i$s can also be expressed in term of $c_z, c_z^\dagger$ 
\begin{align}
&\Gamma_{2i}=\frac{i}{\sqrt{N}}\sum_{z\in S_{-}} (z^ic^{\dagger}_z-z^{-i}c_z) \ ,   \label{eq:mode3}\\
&\Gamma_{2i+1}=-\frac{1}{\sqrt{N}}\sum_{z\in S_{-}} (z^{i}C(\bar z)c_z^{\dagger}+z^{-i}C(z)c_z) \ .  \label{eq:mode4}
\end{align}
To obtain the form factors of the charge odd operators such as the spin operator, one needs to compute the matrix elements between the two Fock basis defined by two sets of creation-annihilation operators~\cite{cmp/1103901557,cmp/1103904079}. In the continuum limit, this is not a well posed problem. But in the finite $2^N$ dimensional Hilbert space, since both of the  $c_z, c_z^\dagger$ and $c_\beta, c_\beta^\dagger$ are ultimately defined linearly by the same sets of $\Gamma_i$s, this problem is well posed and can be reduced to matrix inversion and multiplication problems in the $N\times N$ one-particle Hilbert space. We will show in Sec.~\ref{sec:derivation}, that the matrices to be inverted and multiplied at the critical point are rather special. As a result, elementary derivations of the form factors exist.

Before moving to the form factors, let's stay in this introductory subsection a little bit longer and address the property of the eigensystem Eq.~(\ref{eq:systemNS}) and Eq.~(\ref{eq:systemR}) under the spatial translation. The first observation to make is, although $|\Omega_R\rangle$ is not an eigenstate of the $H$ in Eq.~(\ref{eq:Hamil}), it is the absolute ground state of another ``twisted'' Hamiltonian which relates to $H$ by a sign flip in the $\sigma^x_{N-1}\sigma^x_0$ term
\begin{align}
H'=-\frac{1}{2}\sum_{i=0}^{N-2}\sigma^x_i\sigma^x_{i+1}+\frac{1}{2}\sigma^x_{N-1}\sigma^x_0+\frac{1}{2}\sum_{i=0}^{N-1}\sigma^z_i \ . 
\end{align}
Indeed, the fermion quadratic form for $H'$ simply differs from Eq.~(\ref{eq:fermH}) by a sign flip in the $Z_2$ dependent term. As such, the spectrum for $H'$ can still be expressed in terms of the same sets of $c_z^\dagger, c_\beta^\dagger$s as $H$, but with even numbers of excitations on top of $|\Omega_R\rangle$, and odd numbers of excitations on top of $|\Omega_{NS}\rangle$. We then introduce the following translation operations
\begin{align}
&T_+\sigma^a_{i}T_+^{-1}=\sigma^a_{i+1} \ , 0 \le i\le N-2  \ ; T_+\sigma^a_{N-1}T_+^{-1}=\sigma^a_0 \ , a=x,y, z \ . \\
&T_-\sigma^a_{i}T_-^{-1}=\sigma^a_{i+1} \ , 0 \le i\le N-2  \ ; T_-\sigma^a_{N-1}T_-^{-1}=\sigma^z_0\sigma^a_0\sigma^z_0 \ , a=x,y, z \ .
\end{align}
They can be realized as unitary operators on the $2^N$ dimensional Hilbert space. We chose the normalization of the $T_{\pm}$ such that 
\begin{align}
T_-=-\sigma_0^zT_+ \ . \label{eq:TPmrel}
\end{align}
Clearly, $H$ is invariant under $T_+$, while $H'$ is invariant under $T_-$.  Using the fermionization relations Eq.~(\ref{eq:fermionrel}) and Eq.~(\ref{eq:TPmrel}), one can show that 
\begin{align}
&T_{\pm}\Gamma_{i}=\Gamma_{i+2}T_{\mp}, \  0 \le i\le 2N-3 \ , \  T_{\pm} \Gamma_{2N-2}=\pm \Gamma_0T_{\mp} Z \ , T_{\pm} \Gamma_{2N-1}=\pm \Gamma_1T_{\mp} Z \ . 
\end{align}
Using the mode decompositions in Eq.~(\ref{eq:mode1}) and Eq.~(\ref{eq:mode3}), and the fact that $z^{N}=-1$, $\beta^N=1$, one has~\cite{Iorgov:2010ie}
\begin{align}
T_{\pm}c_z^\dagger \frac{1\mp Z}{2}=z c_z^\dagger T_{\mp} \frac{1\mp Z}{2} \ ,  \  T_{\pm}c_\beta^\dagger \frac{1\pm Z}{2}=\beta c_\beta^\dagger T_{\mp} \frac{1\pm Z}{2} \ . \label{eq:transrela} 
\end{align}
As such, when acting on the states Eq.~(\ref{eq:systemNS}) and Eq.~(\ref{eq:systemR}), by moving the $T_{\pm}$  all the way to the right using Eq.~(\ref{eq:transrela}), one has 
\begin{align}
&T_+\prod_{i=1}^{2k} c_{z_i}^{\dagger}|\Omega_{NS}\rangle=\left(\prod_{i=1}^{2k}z_ic_{z_i}^{\dagger} \right) T_+|\Omega_{NS}\rangle \ , \\
&T_+\prod_{i=1}^{2k'+1} c_{\beta_i}^{\dagger}|\Omega_{R}\rangle=\left(\prod_{i=1}^{2k'+1}\beta_i c_{\beta_i}^{\dagger}\right) T_-|\Omega_{R}\rangle \ .  
\end{align}
However, since the $|\Omega_{{NS}}\rangle$ and $|\Omega_R\rangle$ are the absolute ground states of $H$ and $H'$, they must be invariant under $T_+$ and $T_-$ up to two phases 
\begin{align}
T_+|\Omega_{NS}\rangle=e^{i\theta_+}|\Omega_{NS}\rangle \ ,  \ T_-|\Omega_{R}\rangle=e^{i\theta_-}|\Omega_{R}\rangle \ . \label{eq:traphase}
\end{align}
As such, the eigensystem Eq.~(\ref{eq:systemNS}) and Eq.~(\ref{eq:systemR}) are indeed eigenstates of the spatial translation $T_+$. Moreover, since we have chosen $T_-=-\sigma^z_0T_+$, one can write
\begin{align}
e^{-i(\theta_+-\theta_-)}\langle \Omega_R|\Omega_{NS}\rangle=\langle \Omega_R|T_-T_+^{-1}|\Omega_{NS} \rangle= -\langle \Omega_R|\sigma_0^z|\Omega_{NS}\rangle\ . 
\end{align}
As such, when $\langle \Omega_{R}|\Omega_{NS} \rangle \ne 0$, due to the reality property, the phase difference $e^{-i(\theta_+-\theta_-)}$ must be $\pm1$ and is an absolute quantity independent of the state normalization. Latter we will show explicitly using the overlapping matrix $v_{z\beta}$ define in Eq.~(\ref{eq:solveg2}) that $e^{i(\theta_+-\theta_-)}=1$. Equivalently, the left continuity at $h=1$ and the absence of discontinuities when $h\in [0,1)$ allow the $e^{i(\theta_+-\theta_-)}$ to be computed by the $h=0$ version, which is clearly $1$. Indeed, when $h=0$, $H=H'$, and the absolute ground state is a trivial ``all-down state''.

\subsection{Square-root products and form-factors at finite $N$}
The major focus of the paper is on the matrix elements of the spin-operator between the fermionic states. Due to the property under spatial translations discussed at the end of the previous subsection, it is sufficient to locate the spin operator at the starting point of the fermionization
\begin{align}
\sigma^x_0=\Gamma_0=\frac{i}{\sqrt{N}}\sum_{z\in S_-}(c_z^\dagger-c_z)=\frac{i}{\sqrt{N}}\sum_{\beta\in S_+}(c_\beta^\dagger-c_\beta) \ . 
\end{align}
We will show latter that the ground state overlap $\langle \Omega_R|\Omega_{NS} \rangle$ is non-vanishing and approaches to $0$ slowly as ${\cal O}(N^{-\frac{1}{4}})$ when $N\rightarrow\infty$. Since both $|\Omega_{NS}\rangle$ and $|\Omega_R\rangle$ can be purely real, it is our freedom to choose the signs of the two states such that $\langle \Omega_R|\Omega_{NS} \rangle>0$. The sign rescaling will not change the matrix elements for all the $Z_2$-even operators. In terms of the ground state overlap, it is more convenient to define the rescaled spin operator
\begin{align}
\hat \sigma=\frac{\sigma_0^x}{\langle\Omega_{R} | \Omega_{NS}\rangle} \ . \label{eq:defspinr}
\end{align}
The matrix elements which we will compute are (products are all from left to right)
\begin{align}
\langle \Omega_R|\prod_{i=1}^{2k+1}c_{\beta_i} \hat \sigma\prod_{j=1}^{2l}c_{z_j}^{\dagger}|\Omega_{NS}\rangle\equiv (-1)^l{\cal M}_{2k+1,2l}(\{\beta_i\}; \{z_j\}) \ . \label{eq:defM}
\end{align}
The advantages of using the rescaled spin operator are two folds. First, computation of the rescaled form factors is naturally separated from the computation of the ground-state overlap in our algorithm. Second, it is the rescaled form-factors that allow {\it finite-volume scaling limit}, defined as 
\begin{align}
&N \rightarrow \infty , \beta \rightarrow 1, \   N(\beta-1) \rightarrow 2\pi i  \times n \ , n\in Z \ , \\
&N \rightarrow \infty , z\rightarrow 1, \   N(z-1) \rightarrow 2\pi i  \times m\ ,  \  m\in Z+\frac{1}{2} \ . 
\end{align}
Clearly, the numbers $n \in Z$ and $m \in Z+\frac{1}{2}$ correspond to the quantization levels on a cylinder.  The vacuum overlap, on the other hand,  approaches to zero at a power-law exactly specified by the scaling dimension of the spin operator. We will show that 
\begin{align}
\bigg|\langle \Omega_R|\Omega_{NS}\rangle\bigg|^2=\left(\frac{2\pi }{N}\right)^{\frac{1}{4}}G\left(\frac{1}{2}\right)G\left(\frac{3}{2}\right)\exp \bigg(\frac{-\pi ^2}{384 N^2}+\frac{7 \pi ^4}{92160 N^4}+{\cal O}\left(\frac{\pi^6}{N^6}\right)\bigg)\ , \label{eq:ovlasym}
\end{align}
where $G({\cal Z})$ is the Barnes-$G$ function. The constant $(2\pi)^{\frac{1}{4}}G(\frac{1}{2})G(\frac{3}{2})$ will ensure that the {\it rescaled corrector} can be matched back to the infinite-volume version, at distances that are much larger than $1$, but much smaller than $N$. This is similar to the connecting property of the infinite-volume massive scaling functions~\cite{Wu:1975mw}.  One major difference is: instead of appearing in the short distance expansion of the scaling function, the constant $G(\frac{1}{2})G(\frac{3}{2})$ moves to the ground state overlap, a quantity that resembles more the spontaneous magnetization in the infinite volume version. In fact, we will show that 
\begin{align}
\bigg|\langle \Omega_{R}|c_1\sigma^x_0|\Omega_{NS}\rangle\bigg|=\bigg|\langle\Omega_R| \Omega_{NS}\rangle\bigg| \ .
\end{align}
As such, Eq.~(\ref{eq:ovlasym}) is also a statement concerning the more standard one-point function of the spin-operator.

For the convenience of readers, below we provide a result list for all the computed form-factors. The results at a generic $N$ are given by Eq.~(\ref{eq:formresult}), and their scaling limits are in Eq.~(\ref{eq:formCFT1}), Eq.~(\ref{eq:formCFT2}). It turns out that certain square-root products and their relations are crucial in all the stages of the derivation and appear also in the final results, so we decide to interlude for a while, by introducing first all the required square-root products. In our algorithm, they will be naturally generated after inverting certain Cauchy matrices and performing further summations. For $\beta \in S_+$, one needs two products
\begin{align}
&p(\beta)=\frac{\prod_{\beta'\in S_+}(\sqrt{\beta}+\sqrt{\beta'})}{\prod_{z'\in S-}(\sqrt{\beta}+\sqrt{z'})} \ , \label{eq:deffb} \\
&q(\beta)=\frac{\prod_{z'\in S_-}(\sqrt{\beta}-\sqrt{z'})}{\prod_{\beta'\in S_+, \beta'\ne \beta}(\sqrt{\beta}-\sqrt{\beta'})} \ , \label{eq:defgb}
\end{align}
and similarly for $z\in S_-$
\begin{align}
&p(z)=\frac{\prod_{\beta'\in S_+}(\sqrt{z}+\sqrt{\beta'})}{\prod_{z'\in S-}(\sqrt{z}+\sqrt{z'})} \ , \label{eq:deffz} \\
&q(z)=\frac{\prod_{z'\in S_-,\ z'\ne z}(\sqrt{z}-\sqrt{z'})}{\prod_{\beta'\in S_+}(\sqrt{z}-\sqrt{\beta'})} \ . \label{eq:defgz} 
\end{align}
In these products, we use $\sqrt{1}=1$. They satisfy the simple relations
\begin{align}
q(\beta)=\frac{\sqrt{\beta}}{N}p(\beta) \ , \   q(z)=\frac{N}{\sqrt{z}}p(z) \ , \label{eq:refg}
\end{align}
which allows to use only the products $p(\beta)$ and $p(z)$ in the final results. To show this relation, we introduce the $2N$-th roots of unity\footnote{They are the momenta sets one could encounter, if one decide to treat even and odd sites equally when diagonalizing the rotations. But then, the pairing between the positive and negative energies restrict the excitation momenta to the upper half plane, which are essentially the square-roots of elements in $S_\pm$. }
\begin{align}
S^{2N}_\pm=\{\zeta; \zeta^{2N}=\pm 1\} \ . 
\end{align}
They can be expressed as unions of $\pm \sqrt {z}, \  \pm \sqrt{\beta}$ of elements in $S_{\pm}$
\begin{align}
&S^{2N}_-=\{\sqrt{z}; z\in S_-\}\cup \{-\sqrt{z}; z\in S_-\} \ , \\
&S^{2N}_+=\{\sqrt{\beta}; \beta\in S_+\}\cup \{-\sqrt{\beta}; \beta\in S_+\} \ ,  
\end{align}
where we use $\sqrt{1}=1$ . By expanding the logarithms, it is not hard to show the following 
\begin{align}
\sum_{\zeta\in S_+^{2N}, \  \zeta\ne \sqrt{\beta}}&\ln\left(1-\frac{\zeta}{\sqrt{\beta}}\right)=\lim_{\lambda\rightarrow 1^-}\sum_{\zeta\in S_+^{2N}, \  \zeta\ne 1}\ln \left(1-\lambda\zeta\right)=\ln (2N) \ , \\
\sum_{\zeta\in S_-^{2N}}&\ln\left(1-\frac{\zeta}{\sqrt{\beta}}\right)=\lim_{\lambda\rightarrow 1^-}\sum_{k=1}^{\infty}\frac{(-1)^{k-1}\lambda^{2Nk}}{k}=\ln 2 \ , 
\end{align}
and similar for sums with $\sqrt{z}$ in the denominators. As such, we can write 
\begin{align}
&\frac{p(\beta)}{q(\beta)}=\frac{\prod_{\zeta\in S^{2N}_+, \  \zeta \ne\sqrt{\beta} }(\sqrt{\beta}-\zeta)}{\prod_{\zeta\in S^{2N}_-}(\sqrt{\beta}-\zeta)}=\frac{2N}{2\sqrt{\beta}}=\frac{N}{\sqrt{\beta}} \ , \\
&\frac{p(z)}{q(z)}=\frac{\prod_{\zeta\in S^{2N}_+}(\sqrt{z}-\zeta)}{\prod_{\zeta\in S^{2N}_-, \ \zeta\ne \sqrt{z}}(\sqrt{z}-\zeta)}=\frac{2\sqrt{z}}{2N}=\frac{\sqrt{z}}{N} \ ,
\end{align}
proving the desired relations. We can combine the above with the product formula Eq.~(\ref{eq:prod}) to derive closed expressions at $\beta=1$. In fact,  Eq.~(\ref{eq:prod}) implies that 
\begin{align}
\frac{p(1)q(1)}{2}=\frac{i+1}{i-1}=-i \ . \label{eq:pq1}
\end{align}
As such, to obtain $p(1)$ it remains to know its phase. However, using the paring relation in $S_{\pm}$, and using the fact that in the $(0,2\pi)$ branch one has $\sqrt{z^{-1}}=-\frac{1}{\sqrt{z}}$, for any pair appearing in the product, one has the following contribution with $\alpha>0$
\begin{align}
(1+e^{\frac{i\alpha}{2}})(1-e^{-\frac{i\alpha}{2}})=2i\sin \frac{\alpha}{2} \ . 
\end{align}
Since $\alpha>0$, $p(1)$ is then a positive number multiplying
\begin{align}
(1+i)(i)^{\frac{N-2}{2}}(-i)^{\frac{N}{2}}=1-i \ . 
\end{align}
Combing all above,  one obtains the exact values
\begin{align}
p(1)=\sqrt{N}(1-i), \  q(1)= \frac{1}{\sqrt{N}}(1-i) \ .  \label{eq:exactfg1}
\end{align}
They will be used latter throughout the derivation, in particular, to show a crucial matrix is invertible.  On the other hand, for $\beta\ne 1$, no such simple formulas can be derived for the $p$ function. From the perspective of the $S^{2N}_{\pm}$ circles, this is because all the products are restricted to the upper half-plane. But for both $S_+$ and $S_-$, the $p$ function can still be simplified into a single integral representation. What one needs are the following 
\begin{align}
&p({\cal Z})=\exp \bigg(\hat \Gamma_N({\cal Z})-\frac{1}{2}\ln \left(1-\frac{1}{\sqrt{{\cal Z}}}\right)+\frac{1}{2}\ln \left(1+\frac{1}{\sqrt{{\cal Z}}}\right)\bigg) \ , \label{eq:integralp} \\
& \hat \Gamma_N({\cal Z})=-i\left(\sqrt{{\cal Z}}-\frac{1}{\sqrt{{\cal Z}}}\right)\int_0^1\frac{dt}{2\pi \sqrt{t}}\frac{1+t}{(1-{\cal Z} t)(1-{\cal Z}^{-1}t)}\ln \frac{1-t^N}{1+t^N} \ , \label{eq:defGamma}
\end{align}
where ${\cal Z}$ can be both $\beta \in S_+/\{1\}$ and $z\in S_-$. In the expressions, the logarithms are define with the principal branch.  The function $\hat \Gamma_N({\cal Z})$ is  purely real when ${\cal Z} \in (S_+/\{1\})\cup S_-$, and allows analytic continuation to the whole complex plane, with a power-law bound at infinity, and a branch cut along the positive real axis. 

Given all the square-root products. we now list the form factors.  To further simplify the expressions, we introduce the following dressing functions
\begin{align}
&d(\beta)=\frac{\sqrt{\beta}}{\sqrt{N}}\frac{\hat p(\beta)}{1+\sqrt{\beta}} \ , \   \hat p(\beta)=p(\beta)(1+i\delta_{\beta,1}) \ , \ d(1)=1\ , \label{eq:defdressb} \\ 
&d(z)=\frac{1}{\sqrt{N}(\sqrt{z}-1)p(z)} \ . \label{eq:defdressz} 
\end{align}
In terms of the dressing functions, the form factors define in Eq.~(\ref{eq:defM}) read 
\begin{align}
&{\cal M}_{2k+1,2l}(\{\beta_i\};\{z_j\})=i^{k+l+1}\prod_{i<i'}\frac{\sqrt{\beta_i}-\sqrt{\beta_j}}
{\sqrt{\beta_i}\sqrt{\beta_j}-1}\prod_{i,j}\frac{\sqrt{\beta_i}\sqrt{z_j}-1}{\sqrt{\beta_i}-\sqrt{z_j}}\prod_{j<j'}\frac{\sqrt{z_j}-\sqrt{z_{j'}}}{\sqrt{z_j}\sqrt{z_{j'}}-1} \nonumber \\ 
&\times \prod_{i=1}^{2k+1} d(\beta_i)\prod_{j=1}^{2l} d(z_j) \  . \label{eq:formresult}
\end{align}
Notice that within $\{\beta_i\}$ at most there can be only one $\beta =1$, and all the denominators are well defined. This is the first major result of the work. Furthermore, in terms of the $\hat \Gamma_N$ function, the ground state overlap has a simple representation
\begin{align}
\bigg|\langle \Omega_R|\Omega_{NS}\rangle\bigg|^2&=\left(\frac{2}{N}\right)^{\frac{1}{4}}\exp \left(\frac{1}{2}\sum_{z\in S_-}\hat \Gamma_N(z)-\frac{1}{2}\sum_{\beta \in S_+, \  \beta\ne 1}\hat \Gamma_N(\beta)\right) \ , \label{eq:overlap}  \\
&\equiv \left(\frac{2}{N}\right)^{\frac{1}{4}}\exp\bigg(\int_0^1 dt \frac{(1-\sqrt{t}) (t^N+\sqrt{t})}{(1+\sqrt{t})^2  \left(t^N+1\right)}\frac{1}{2t\ln t}\bigg) \ . \label{eq:intlap}
\end{align}
Notice in Eq.~(\ref{eq:intlap}), the integrand has a $\ln t$ in the denominator. It is this logarithm that generates the Barnes-$G$ functions in the large $N$ asymptotics Eq.~(\ref{eq:ovlasym}). 

Here we prove the integral representation for $p(\beta)$, $p(z)$ given in Eq.~(\ref{eq:integralp}) and Eq.~(\ref{eq:defGamma}). We first consider the sum with $\beta \in S_{+}/\{1\}$
\begin{align}
SL_+(\beta)=\sum_{\beta'\in S_+,  \ \beta'\ne 1} \ln \left(1+\frac{\sqrt{\beta'}}{\sqrt{\beta}}\right) \ .
\end{align}
Here, the logarithm is in the principal branch and all the square roots are in the $(0,2\pi)$ branch. As an analytic function of $\beta'$, there is a branch cut singularity along the positive real axis, but no other singularities when $\beta'$ is away from the positive real axis. As such, one can introduce the following contour representation 
\begin{align}
SL_+(\beta)=\oint_{\cal C_+}\frac{dw}{2\pi i w} \frac{N}{w^N-1} \ln \left(1+\frac{\sqrt{w}}{\sqrt{\beta}}\right) \ . 
\end{align}
The contour ${\cal C}_+$ is a union of four segments
\begin{align}
{\cal C}_+=&\bigg\{0.9e^{i\theta}, -\frac{\pi}{N}\le\theta\le\frac{\pi}{N} \bigg\}\bigcup\bigg\{1.1e^{i\theta}, -\frac{\pi}{N}\le\theta\le\frac{\pi}{N} \bigg\}\nonumber \\ &\bigcup \bigg\{re^{\frac{i\pi}{N}}, 0.9\le r\le1.1 \bigg\}\bigcup \bigg\{re^{\frac{-i\pi}{N}}, 0.9\le r\le1.1 \bigg\} \ . 
\end{align}
It encircles all the zeros of $w^N-1=0$ counter-clockwisely, except $w=1$. Now, we deform the contour to the branch cut along the positive real axis. The decay at infinity and around $w=0$ are guaranteed, but the $w=1$ pole of the $w^N-1$ needs to be taken care of. Since the upper and lower limits of the logarithm are continuous at $w=1$, one simply needs to pick up half of the upper and lower residues  
\begin{align}
&SL_+(\beta)={\rm PV}\int_0^{\infty} \frac{dt}{2\pi i t } \frac{N}{t^N-1}\ln \left(1+\frac{\sqrt{t}}{\sqrt{\beta}}\right)-{\rm PV}\int_0^{\infty} \frac{dt}{2\pi i t } \frac{N}{t^N-1}\ln \left(1-\frac{\sqrt{t}}{\sqrt{\beta}}\right)  \nonumber \\ 
&-\frac{1}{2}\ln \left(1-\frac{1}{\sqrt{\beta}}\right)-\frac{1}{2}\ln \left(1+\frac{1}{\sqrt{\beta}}\right) \ ,
\end{align}
where $\sqrt{t}>0$, and the last line is due to the half residues. The ${\rm PV}$ are for the singularity at $t=1$. We further simplify the two integrals by splitting the domain into the $(0,1)$ and $(1,\infty)$, and then change the variable according to $t \rightarrow t^{-1}$ in the $(1,\infty)$ part. This leads to
\begin{align}
&{\rm PV}\int_0^{\infty} \frac{dt}{2\pi i t } \frac{N}{t^N-1}\ln \left(1\pm \frac{\sqrt{t}}{\sqrt{\beta}}\right)=-N\int_0^1\frac{dt}{2\pi i  t}\ln \left(1\pm\frac{\sqrt{t}}{\sqrt{\beta}}\right) \nonumber \\ 
&+\int_0^1 \frac{dt}{2\pi i} \frac{N t^{N-1}}{t^N-1} \left(\ln \left(1\pm \frac{\sqrt{t}}{\sqrt{\beta}}\right)-\ln \left(1\pm \frac{1}{\sqrt{t} \sqrt{\beta}}\right)\right) \ . 
\end{align}
We now use the fact that 
\begin{align}
\frac{d}{dt}\ln (1-t^N)=\frac{Nt^{N-1}}{t^{N}-1} \ , 
\end{align}
and partial integrate. After simplifying the resulting integrands, one obtains
\begin{align}
&SL_+(\beta)=\left(\sqrt{\beta}-\frac{1}{\sqrt{\beta}}\right)\int_0^1\frac{dt}{2\pi i \sqrt{t}} \frac{1+t}{(1-\beta t)(1-\bar \beta t)}\ln (1-t^N) -\frac{1}{2}\ln \left(1-\frac{1}{\sqrt{\beta}}\right)\nonumber \\ 
&-\frac{1}{2}\ln \left(1+\frac{1}{\sqrt{\beta}}\right)-N\int_0^1\frac{dt}{2\pi i t}\ln \left(1+\frac{\sqrt{t}}{\sqrt{\beta}}\right)+N\int_0^1\frac{dt}{2\pi i t}\ln \left(1-\frac{\sqrt{t}}{\sqrt{\beta}}\right) \ . \label{eq:SLp}
\end{align}
Similarly, for the sum ($\beta \in S_{+}/\{1\}$)
\begin{align}
SL_-(\beta)=\sum_{z\in S_-}\ln \left(1+\frac{\sqrt{z}}{\sqrt{\beta}}\right) \ , 
\end{align}
we can write it as
\begin{align}
SL_-(\beta)=\int_0^{\infty} \frac{dt}{2\pi i t} \frac{-N}{t^N+1}\left(\ln \left(1+\frac{\sqrt{t}}{\sqrt{\beta}}\right)-\ln \left(1-\frac{\sqrt{t}}{\sqrt{\beta}}\right)\right) \ .
\end{align}
Again split the domain and change $t$ to $t^{-1}$ in the $(1,\infty)$ part, one has 
\begin{align}
&SL_-(\beta)=-N\int_0^1\frac{dt}{2\pi i t}\ln\left(1+ \frac{\sqrt{t}}{\sqrt{\beta}}\right)+N\int_0^1\frac{dt}{2\pi i t}\ln\left(1- \frac{\sqrt{t}}{\sqrt{\beta}}\right) \nonumber \\ 
&+\int_0^1\frac{dt}{2\pi i }\frac{Nt^{N-1}}{t^{N}+1}\bigg(\ln \left(1+\frac{\sqrt{t}}{\sqrt{\beta}}\right)-\ln \left(1+\frac{1}{\sqrt{t}\sqrt{\beta}}\right)-\left(\sqrt{\beta}\rightarrow-\sqrt{\beta}\right)\bigg) \ . 
\end{align}
Using 
\begin{align}
\frac{d}{dt}\ln (1+t^N)=\frac{Nt^{N-1}}{1+t^{N}} \ ,
\end{align}
and partial integrate again, one has 
\begin{align}
&SL_-(\beta)=\left(\sqrt{\beta}-\frac{1}{\sqrt{\beta}}\right)\int_0^1\frac{dt}{2\pi i \sqrt{t}} \frac{1+t}{(1-\beta t)(1-\bar \beta t)}\ln (1+t^N)\nonumber \\ 
&-N\int_0^1\frac{dt}{2\pi i t}\ln \left(1+\frac{\sqrt{t}}{\sqrt{\beta}}\right)+N\int_0^1\frac{dt}{2\pi i t}\ln \left(1-\frac{\sqrt{t}}{\sqrt{\beta}}\right) \ . \label{eq:SLm}
\end{align}
Now, combining the Eq.~(\ref{eq:SLp}), Eq.~(\ref{eq:SLm}) with the definitions Eq.~(\ref{eq:deffb}) and Eq.~(\ref{eq:defGamma}), one obtains
\begin{align}
&p(\beta)=\exp \bigg(SL_+(\beta)-SL_-(\beta)+\ln \left(1+\frac{1}{\sqrt{\beta}}\right)\bigg) \nonumber \\ 
&=\exp \bigg(\hat \Gamma_N(\beta)-\frac{1}{2}\ln \left(1-\frac{1}{\sqrt{\beta}}\right)+\frac{1}{2}\ln \left(1+\frac{1}{\sqrt{\beta}}\right) \bigg) \ , 
\end{align}
which is the desired Eq.~(\ref{eq:integralp}) for ${\cal Z}=\beta\in S_+/\{1\}$. Notice that the extra $\ln (1+\frac{1}{\sqrt{\beta}})$ is due to the $\beta'=1$ term in the Eq.~(\ref{eq:deffb}). The case for ${\cal Z}=z\in S_-$ can be established exactly the same way.

\subsection{Scaling limits of the form factors}
The integral representations allow to take the scaling limit of the form factors. The crucical quantity is the function $\hat \Gamma_N({\cal Z})$ defined in Eq.~(\ref{eq:defGamma}). To obtain its scaling limit, one changes the variable as $t \rightarrow e^{-\frac{t}{N}}$, leading to ($\alpha\in R_{\ne 0}$)
\begin{align}
&\hat \Gamma_N(e^{i\frac{\alpha}{N}})=\sin \frac{|\alpha|}{2N}\int_0^{\infty}\frac{e^{-\frac{t}{2N}}}{\pi N} \ln \frac{1-e^{-t}}{1+e^{-t}} \frac{1+e^\frac{-t}{N}}{(1-e^{\frac{i\alpha-t}{N}})(1-e^{\frac{-i\alpha-t}{N}})}dt \nonumber \\ 
& \rightarrow \int_0^{\infty} \frac{dt}{\pi } \frac{|\alpha|}{t^2+\alpha^2}\ln \frac{1-e^{-t}}{1+e^{-t}}-\frac{\pi|\alpha|}{96N^2}+{\cal O}\left(\frac{\pi|\alpha|^3}{N^4}\right) \ .  \label{eq:Gammalim}
\end{align}
Clearly, this is because that when ${\cal Z} -1={\cal O}(N^{-1})$, the integral in Eq.~(\ref{eq:defGamma}) receives the largest contribution only in the scaling region when $t-1={\cal O}(N^{-1})$. Here the crucial integral is 
\begin{align}
 \mathfrak{b}(\alpha)=\int_0^{\infty} \frac{dt}{\pi } \frac{\alpha}{t^2+\alpha^2}\ln \frac{1-e^{-t}}{1+e^{-t}} \equiv \ln \sqrt{\alpha}+\ln \frac{\Gamma \left(\frac{\alpha}{2\pi}+\frac{1}{2}\right)}{\sqrt{2\pi}\Gamma(\frac{\alpha}{2\pi}+1)} \ ,  \ \alpha>0 \ . \label{eq:binet}
\end{align}
This is a combination of the Binet's second integral and its generalization by Hermite~\cite{Hermite+1895+201+208,Boyack+2023}. To convince oneself this non-trivial formula, one can use the following Barnes representation
\begin{align}
\mathfrak{b}(\alpha)=-\int_{0<\rm Re (u)<1} \frac{du}{2\pi i} \left(1-\frac{1}{2^{u+1}}\right) \frac{ \Gamma (u) \zeta(u+1)}{\cos \frac{\pi u}{2}} \alpha^{-u} \ , \label{eq:Barnb}
\end{align}
which allows to compute the small $\alpha$ expansion of $\mathfrak{b}(\alpha)$ and compare with that of the $\Gamma$ function ratios. Notice that throughout this work, the contours for such Barnes integrals are always along the imaginary direction from ${\rm Re}(u)-i\infty$ to ${\rm Re}(u)+i\infty$.

Given the scaling limit of the $\hat \Gamma_N$, the scaling limits of the dressing functions $d(\beta)$ and $d(z)$ naturally follow. Here we introduce the following dressing function in the scaling limit
\begin{align}
\tau(n)=\frac{\Gamma(n+\frac{1}{2})}{\sqrt{2\pi}\Gamma(n+1)} \ , \label{eq:deftau}
\end{align}
and further separate the non-zero quantization levels in the scaling limit to the left ($L$) and right ($R$) sectors
\begin{align}
&L: \beta_n^+=e^{i\frac{2\pi n}{N}} ;  \  R:  \beta_n^-=e^{-i\frac{2\pi  n}{N}}  \ , \ n\in Z_{\ge 1} \ , \\
&L: z_m^+=e^{i\frac{2\pi m}{N}}; \ R:  z_m^-=e^{-i\frac{2\pi m}{N}} \ , \  m\in Z_{\ge 0}+\frac{1}{2} \ . 
\end{align}
Then, the dressing functions approach the following limits
\begin{align}
&d(\beta_n^+) \rightarrow e^{-\frac{i\pi}{4}}\tau(n) \ , \  d(\beta_n^-) \rightarrow e^{i\frac{\pi }{4}} \tau(n) \ , \\ 
&d(z_m^+)\rightarrow e^{-i\frac{\pi}{4}}\tau\left(m-\frac{1}{2}\right) \ ,  \ d(z_m^-)\rightarrow -e^{i\frac{\pi}{4}}\tau\left(m-\frac{1}{2}\right) \ . 
\end{align}
Notice that at the zero momentum, $d(1)=1$ is a constant. In term of the above, the form factors in the scaling limit can be further simplified.

To present the results, we first clarify the conventions we used in the scaling limit. In all the matrix elements, we label the creation and annihilation operators from the left to the right. On each sides of the spin operator, the plus momenta (L) are always placed left to all the minus momenta (R).  Without zero mode, we use the integer sequence $(n_L,n_R;m_L,m_R)$ in the subscript of the matrix element to denote that there are $n_L$ $\beta^+$s, $n_R$ $\beta^-$s, and $m_L$ $z^+$s, $m_R$ $z^-$s. 
And $(0,n_L,n_R;n_L,n_R)$ denotes that in addition there is a zero mode $c_1$, which is always placed next to the $\langle \Omega_R|$. The quantization levels for these modes are collected by four ordered sets $N_L, N_R, M_L, M_R$, where $N_L, N_R$ consist of positive integers, and $M_L$, $M_R$ consist of positive half integers. To summarize, in the scaling limit, the momenta in the matrix elements are collected by the following ordered sets
\begin{align}
&\{\beta_i\}_{(N_L,N_R)}= \{\beta_{n}^+, n\in N_L\}\bigcup \{\beta_{n}^-, n\in N_R\} \  ; \  |N_L|=n_L, \ |N_R|=n_R  \ ,  \\ 
&\{z_j\}_{(M_L,M_R)}=\{z_{m}^+, m\in M_L\}\bigcup \{z_{m}^-, m\in M_R\}  \  ; \  |M_L|=m_L, \ |M_R|=m_R  \ ,\\ 
& \{\beta_i\}_{(0, N_L,N_R)}=\{1\}\bigcup\{\beta_i\}_{(N_L,N_R)} \ , 
\end{align}
where the orderings are always from the left to the right, and $|A|$ denotes the size of the set $A$. To present the scaling limit, we further introduce the following products
\begin{align} 
& D(N)=\prod_{n\in N}\tau(n) \ ,  \ D(M)=\prod_{m\in M}\tau\left(m-\frac{1}{2}\right) \ , \ \Pi\left(N,M\right)=\prod_{n\in N,m\in M}\frac{n+m}{n-m}  \ , \nonumber \\ 
&\Pi\left(N\right)=\prod_{n_i, n_j\in N , \ i<j}\frac{n_i-n_{j}}{n_i+n_{j}} \ , \  \Pi\left(M\right)=\prod_{m_i, m_j\in M , \ i<j}\frac{m_i-m_{j}}{m_i+m_{j}}  \ , 
\end{align}
where $N$ is a set positive integers, and $M$ is a set of positive half integers. A product equals to $1$ when one or both of the sets appearing in the product are empty. In terms of the products, scaling limits of the matrix elements without zero mode are given by
\begin{align}
&{\cal M}_{n_L+n_R; \ m_L+m_R}\left(N_L,N_R; M_L,M_R\right)=Z_{n_L,n_R;m_L,m_R} D(N_L)D(N_R)D(M_L)D(M_R)\nonumber \\ 
&\times \Pi(N_L)\Pi(N_R)\Pi(M_L)\Pi(M_R)\Pi(N_L,M_L)\Pi(N_R,M_R) \ , \label{eq:formCFT1}
\end{align}
and with a zero mode one has 
\begin{align}
&{\cal M}_{n_L+n_R+1; \ m_L+m_R}\left(0, N_L,N_R; M_L,M_R\right)=Z_{0, n_L,n_R;m_L,m_R} D(N_L)D(N_R)D(M_L)D(M_R)\nonumber \\ 
&\times \Pi(N_L)\Pi(N_R)\Pi(M_L)\Pi(M_R)\Pi(N_L,M_L)\Pi(N_R,M_R) \ . \label{eq:formCFT2}
\end{align}
In Eq.~(\ref{eq:formCFT1}), the excitation numbers are subjected to the restriction $n_L+n_R=1 \mod 2$, $m_L+m_R=0\mod 2$. And in Eq.~(\ref{eq:formCFT2}) one has $n_L+n_R=0\mod 2$, $m_L+m_R=0\mod 2$. The phase factors depend only on the even-odd parities of these excitation numbers. If we use $+$ to denote the even parity, and $-$ to denote the odd parity, then the phase factors read 
\begin{align}
&Z_{-, +;+,+}=Z_{+,-;-,-}=e^{i\frac{\pi}{4}} \ ,  \ Z_{-,+;-,-}=-Z_{+,-;+,+}=e^{-i\frac{\pi}{4}} \ ; \nonumber \\ 
&Z_{0,+,+;+,+}=Z_{0,-,-;-,-}=i \ , \ Z_{0,-, -; +, +}=-Z_{0, +, +; - ,-}=1 \ .  \label{eq:phases}
\end{align}
To obtain the phase factors, one simply notice that in the scaling limit, the two particle functions in the R sector simplify as (formulas with in the NS sector are similar)
\begin{align}
&\frac{\sqrt{\beta^\sigma_n}-\sqrt{\beta_{n'}^\sigma}}{\sqrt{\beta_n^\sigma}\sqrt{\beta_{n'}^\sigma}-1} \rightarrow \sigma \frac{n-n'}{n+n'} \ , \frac{\sqrt{\beta^\sigma_n}-\sqrt{\beta_{n'}^{-\sigma}}}{\sqrt{\beta_n^{\sigma}}\sqrt{\beta_{n'}^{-\sigma}}-1} \rightarrow -\sigma  \ , \  \sigma =\pm \ . 
\end{align}
Between the two sectors, instead one has  
\begin{align}
&\frac{\sqrt{\beta_n^\sigma}\sqrt{z_m^\sigma}-1}{\sqrt{\beta_n^\sigma}-\sqrt{z_m^\sigma}} \rightarrow \sigma \frac{n+m}{n-m} \ ,  \ \frac{\sqrt{\beta_n^\sigma}\sqrt{z_m^{-\sigma}}-1}{\sqrt{\beta_n^\sigma}-\sqrt{z_m^{-\sigma}}} \rightarrow -\sigma 
\ . 
\end{align}
For the two particle functions involving $\beta=1$, simply notice that in our ordering they all become $-1$. Collecting all the $-1$ and the $e^{\pm \frac{i\pi}{4}}$ factors leads to the final expressions Eq.~(\ref{eq:phases}) for the phase factors.  The explicit expressions Eq.~(\ref{eq:formCFT1}), Eq.~(\ref{eq:formCFT2}) and Eq.~(\ref{eq:phases}) for the form factors in the scaling limit
are the second major result of the paper. They provide explicit formulas for the CFT matrix elements computed recursively in~\cite{Yurov:1991my}, up to the phase conventions.
Notice that all the form factors in the scaling limit are rational numbers in our convention, due to the overall $e^{\pm \frac{i\pi}{4}}$ 
in the phase factors. 

Given the matrix elements in the fermionic basis, one can establish the form factor expansions for the multi-point correlators. We will show in Sec.~\ref{sec:scale} that the form factor expansion of the rescaled two-point correlator in the $NS$ ground state can be summed into a product of two Fredholm determinants of the form Eq.~(\ref{eq:abs}) given in the abstract, one for the left sector, another for the right sector. Assuming the identity $\det (1+{\cal K})=(1-w)^{-\frac{1}{8}}$ checked up to $w^{30}$ in the small-$w$ expansion, this leads exactly to the standard CFT 2pt on a cylinder. This is another major result of the work.

\section{Derivation of the main results} \label{sec:derivation}

After introducing all the main results, in this section we derive the results presented in the previous sections. The overall strategy is similar to~\cite{cmp/1103901557,cmp/1103904079}\footnote{This strategy is essentially equivalent to that of~\cite{Iorgov:2010ie}, but allows to avoid the explicit Gaussian wave functions of the ground states. }, but the way of solving the matrix equation is different. Rather than using the Wiener-Hopf trick suitable for the infinite-volume limit, our approach is based on the near-Cauchy property of the $W$, $W'$ matrices given below in Eq.~(\ref{eq:defW}). This special property of the critical model allows to obtain explicit product formulas at finite $N$ using relatively elementary methods, bypassing the elliptic Cauchy determinant identities~\cite{Iorgov:2010ie} required in more general cases. On the other hand, $W$ and $W'$
are Cauchy only up to rank-one perturbations at $\beta=1$, due to the subtle behavior of the critical model at the zero momentum. Not only the square-root function $\varphi(\beta)=\sqrt{\beta}$ for $\beta \in S_+/\{1\}$ has a branch cut along the positive real axis, but also, $\varphi(1)=i$ can never be the upper or lower limits of the square-root at $\beta=1$:  otherwise the first column of the matrix $\Omega$ in Eq.~(\ref{eq:rotat1}), Eq.~(\ref{eq:rotat2}) would vanish, and $\Omega$ would no-longer be a rotation. This is another special feature of the critical model.  

Our derivation can be separated into three stages. First, we derive the crucial {\it overlapping matrices} between the $NS$ and $R$ ground states. There are three types of them
\begin{align}
&\frac{\langle \Omega_R|c_\beta c_{\beta'}|\Omega_{NS}\rangle}{\langle \Omega_R|\Omega_{NS}\rangle}=u(\beta,\beta') \ , \label{eq:defu1} \\
&\frac{\langle \Omega_R|c_z^\dagger c_{z'}^{\dagger}|\Omega_{NS}\rangle}{\langle \Omega_R|\Omega_{NS}\rangle}=u(z,z') \ , \label{eq:defu2} \\
&\frac{\langle \Omega_R|c_\beta c_{z}^\dagger|\Omega_{NS}\rangle}{\langle \Omega_R|\Omega_{NS}\rangle}=u(\beta,z) \ . \label{eq:defu3}
\end{align}
Since both of the $|\Omega_R\rangle $ and  $|\Omega_{NS}\rangle$ are Gaussian-type ground states of fermionic quadratic forms defined in terms of the same set of fermionic operators, multi-point correlators between the two states are all generated by the overlapping matrices through Wick-contractions~\cite{Perk:1983hg}. In particular, since the $\sigma^x_0=\Gamma_0$ is also a fermionic operator, matrix elements with $\Gamma_0$ can also be obtained from Wick contractions.  It turns out that it is more convenient to introduce the following matrices
\begin{align}
g_{\beta,\beta'}=u(\beta,\bar \beta'),  \ \hat g_{z,z'}(z,z')=u(z,\bar z') \ , \  v_{z\beta}=u(\beta,z) \ .  \label{eq:defg}
\end{align}
In terms of which, one has the following contractions with $\Gamma_0$
\begin{align}
\langle c_\beta\Gamma_0\rangle=\frac{i}{\sqrt{N}}\left(1-\sum_{\beta'}g_{\beta \beta'}\right) \ , \\
\langle \Gamma_0c_z^\dagger \rangle=\frac{i}{\sqrt{N}}\left(-1+\sum_{z'}\hat g_{z' \bar z}\right) \ , 
\end{align}
where we have used the mode decompositions in Eq.~(\ref{eq:mode1}) and Eq.~(\ref{eq:mode3}). As such, given the overlapping matrices and the contractions with $\Gamma_0$, in the second stage we perform the Wick contraction to obtain the result Eq.~(\ref{eq:formresult}) of the form factors. Finally, in the third stage, we derive the Eq.~(\ref{eq:overlap}) for the ground state overlap  $|\langle\Omega_R |\Omega_{NS}\rangle|$ as an application of the form factors we derived. For notation simplicity, below we will use following notation for the average 
\begin{align}
\langle O\rangle \equiv \frac{\langle \Omega_R|O|\Omega_{NS}\rangle}{\langle \Omega_R|\Omega_{NS}\rangle} \ , 
\end{align}
where $O$ can be a generic operator.

To proceed, we first derive the equations satisfied by the matrices $g, \hat g, v$ in Eq.~(\ref{eq:defg}). For this, we first notice that the formula 
\begin{align}
\sum_{i=0}^{N-1}(\beta/z)^i \equiv \frac{2}{1-\frac{\beta }{z}} \ , \beta\in S_+, z\in S_- \ , \label{eq:sumbz}
\end{align}
allows to convert between $c_z$ and $c_\beta$
\begin{align}
&c_z^{\dagger}-c_{\bar z}=\frac{2}{N}\sum_{\beta \in S_+}\frac{c^\dagger_\beta-c_{\bar \beta}}{1-\frac{\beta}{z}} \, \\ 
&c_z^{\dagger}+c_{\bar z}=\frac{2}{N}\sum_{\beta \in S_+}\frac{c_\beta^\dagger+c_{\bar \beta}}{1-\frac{\beta}{z}} \frac{C(z)}{C(\beta)}   \ .
\end{align}
Similarly, one can also express $c_\beta$, $c_\beta^\dagger$ in terms of $c_z, c_z^\dagger$ as 
\begin{align}
c_\beta^{\dagger}-c_{\bar \beta}=\frac{2}{N}\sum_{z\in S_-}\frac{c_z^{\dagger}-c_{\bar z}}{1-\frac{z}{\beta}} \ , \\
c_\beta^{\dagger}+c_{\bar \beta}=\frac{2}{N}\sum_{z\in S_-}\frac{c_z^{\dagger}+c_{\bar z}}{1-\frac{z}{\beta}} \frac{C(\beta)}{C(z)} \ . \label{eq:cover2}
\end{align}
To show the above, we start from the mode decompositions Eq.~(\ref{eq:mode3}) and Eq.~(\ref{eq:mode4}). We can use the relations Eq.~(\ref{eq:gammab}), Eq.~(\ref{eq:gammbhf}) and Eq.~(\ref{eq:cb}), or equivalently, the standard summation formula
\begin{align}
\sum_{i=0}^{N-1}(z\bar z')^i=N\delta_{zz'} \ , z, z' \in S_- \ , 
\end{align}
and the relation $C(z)C(\bar z)=C(\beta)C(\bar \beta)=1$ to express $c_z$ or $c_\beta$ in terms of $\Gamma_i$s 
\begin{align}
c_z^{\dagger}-c_{\bar z}=\frac{1}{i\sqrt{N}}\sum_{i=0}^{N-1}\Gamma_{2i}z^{-i} \ ,  \ 
c_z^{\dagger}+c_{\bar z}=-\frac{1}{\sqrt{N}}\sum_{i=0}^{N-1}\Gamma_{2i+1}C(z)z^{-i} \ . \label{eq:convertc}
\end{align}
Then, we express $\Gamma_i$ in terms of $c_\beta, c_\beta^\dagger$ using Eq.~(\ref{eq:mode1}), Eq.~(\ref{eq:mode2}) and use the sum rule Eq.~(\ref{eq:sumbz}) to simplify the results.

Knowing the conversion rules, we now consider the following identity~\cite{cmp/1103901557}
\begin{align}
\langle \Omega_R|c_{\beta}c_{\bar z}|\Omega_{NS}\rangle \equiv 0 \ .
\end{align}
Now, express the $c_{\bar z}$ using $c_\beta$ and $c_{\beta}^{\dagger}$, and then move the $c_\beta^{\dagger}$s to the left, one obtains
\begin{align}
\sum_{\beta'\in S_+} \langle\Omega_R| c_\beta c_{  \bar \beta'}|\Omega_{NS}\rangle\frac{C(\bar \beta')+C(\bar z)}{\beta'-z} =-\frac{C\left(\bar \beta\right)- C(\bar z)}{\beta-z} \langle \Omega_R |\Omega_{NS}\rangle \ . \label{eq:mrtemp}
\end{align}
At this stage, it is time to introduce the following $N\times N$ matrices 
\begin{align}
W=W_{\beta z}=\frac{\varphi(\beta)+\varphi(z)}{\beta-z} \ ,  \ W'=W'_{\beta z}=-\frac{\varphi(\beta)-\varphi(z)}{\beta-z} \ , \label{eq:defW}
\end{align}
where $C(\bar {\cal Z})=-i\varphi({\cal Z})$. Here we show that, if $W$ is invertible, then the ground state overlap $\langle \Omega_R |\Omega_{NS}\rangle $ must be non-vanishing. Assume the opposite, then from Eq.~(\ref{eq:mrtemp}), the overlapping matrix with two $c_\beta$s must be identical to zero. Then, using the same trick for 
\begin{align}
\langle \Omega_{R}|c_{\beta_1}c_{\beta_2}c_{\beta_3}c_{\bar z}|\Omega_{NS} \rangle=0 \ , 
\end{align}
and move the $c_{\beta}^\dagger$s in $c_{\bar z}$ to the left, one can show that all the $\langle \Omega_{R}|c_{\beta_1}c_{\beta_2}c_{\beta_3}c_{\beta_4}|\Omega_{NS} \rangle$ are also zero. Recursively\footnote{This method can also be used to prove the Wick theorem in the current context, as it lead to recursive relations for Pfaffians~\cite{cmp/1103901557}.}, this implies that all the matrix elements $\langle \Omega_{R}|c_{\beta_1}c_{\beta_2}c_{\beta_3}...c_{\beta_{2n}}|\Omega_{NS} \rangle$ are zero. But this is impossible, because $c_{\beta_1}^\dagger c^{\dagger}_{\beta_2}...c^{\dagger}_{\beta_n}|\Omega_R\rangle$  is a complete set of orthogonal basis of the full $2^N$ dimensional Hilbert space, and vanishing of all the $\langle\Omega_{R}|c_{\beta_1}c_{\beta_2}c_{\beta_3}...c_{\beta_{2n}}|\Omega_{NS} \rangle$ (odd numbers naturally vanish due to the charge condition) would imply that $|\Omega_{NS}\rangle \equiv 0$. We will show below that $W$ is indeed invertible. As such, by dividing the $\langle \Omega_R |\Omega_{NS}\rangle$ on the two sides of Eq.~(\ref{eq:mrtemp}) one obtains the matrix equation
\begin{align}
gW=W' \rightarrow g=W'W^{-1} \ ,  \label{eq:solveg1}
\end{align}
where $g$ is defined in Eq.~(\ref{eq:defg}). Similarly, we can use Eq.~(\ref{eq:cover2}) in the relations $\langle c_\beta^\dagger c_z^\dagger\rangle= 0$ and $\langle c_{\beta'} c_{\beta}^\dagger \rangle=\delta_{\beta' \beta}$ to derive the following equations
\begin{align}
\hat g=-W^{-1}W' \ ,  \  v_{z\beta}=(W^{-1}){z\beta} \frac{N\varphi_\beta}{\beta} \ , \label{eq:solveg2}
\end{align}
where $\hat g$ and $v$ are defined also in Eq.~(\ref{eq:defg}). The Eq.~(\ref{eq:solveg1}) and Eq.~(\ref{eq:solveg2}) will be the starting point of the following discussions. 

\subsection{Inversion and summation}
Up to now, the discussions are very general and apply to the Ising chain in the whole disordered region, with the $h$-dependent $\varphi$ functions in Eq.~(\ref{eq:defphig}). At the critical point, however, there is a  major simplification: the matrices $W$ and $W'$ are nearly Cauchy, up to rank-one perturbations. This allows to obtain explicit product formulas at a generic $N$ using elementary methods. 

Indeed, using the fact that $\varphi({\cal Z})=\sqrt{{\cal Z}}$ when ${\cal Z} \in (S_{+ }/\{1\}) \cup\ S_-$, and $\varphi(1)=i$, it is easy to see that one has 
\begin{align}
&W_{\beta z}=\frac{\sqrt{\beta}+\sqrt{z}}{\beta-z}+\delta_{\beta1}\frac{i-1}{1-z}\equiv\frac{1}{\sqrt{\beta}-\sqrt{z}}+|0\rangle_\beta\langle v|_z \ , \label{eq: Wdecompose} \\
&W'_{\beta z}=-\frac{\sqrt{\beta}-\sqrt{z}}{\beta-z}-\delta_{\beta1}\frac{i-1}{1-z}\equiv\frac{1}{-\sqrt{\beta}-\sqrt{z}}-|0\rangle_\beta \langle v|_z \ , 
\end{align}
where $|0\rangle$ and $|v\rangle$ are following vectors
\begin{align}
|0\rangle_{\beta}=\delta_{\beta,1} \ , \ \langle v|_z=|v\rangle_z=v_z=\frac{i-1}{1-z}=\frac{i-1}{2}\bigg(\frac{1}{1-\sqrt{z}}-\frac{1}{-1-\sqrt{z}}\bigg) \ . 
\end{align}
Notice that no complex conjugation was performed here, $\langle v|=v^T$ means the transpose of $v$.  As such, we can introduce the following $N\times N$ Cauchy matrices  ($H$ should not be confused with the Hamiltonian)
\begin{align}
T_{\beta z}=\frac{1}{\sqrt{\beta}-\sqrt{z}} \ ,  \  
H_{\beta z}=\frac{1}{-\sqrt{\beta}-\sqrt{z}} \ , \label{eq:defTH}
\end{align}
in terms of which one has 
\begin{align}
W=T+|0\rangle \langle v|  \ ,  \ W'=H-|0\rangle \langle v| \ , v_z=\frac{i-1}{2}\left(T_{1z}-H_{1z}\right) \ .  \label{eq:relWT}
\end{align}
As a Cauchy matrix, $T$ is clearly invertible, as the elements labeling the columns and rows are non-coinciding. Using the explicit product formula for its inverse, one has 
\begin{align}
T^{-1}_{z\beta}&=\frac{1}{\sqrt{\beta}-\sqrt{z}}\frac{\prod_{z'\in S_-}(\sqrt{\beta}-\sqrt{z'})\prod_{\beta'\in S_+}(-\sqrt{z}+\sqrt{\beta'})}{\prod_{\beta'\in S_+, \beta' \ne \beta}(\sqrt{\beta}-\sqrt{\beta'})\prod_{z'\in S_-, z'\ne z}(-\sqrt{z}+\sqrt{z'})}\nonumber \\ 
&=-\frac{1}{\sqrt{\beta}-\sqrt{z}}\frac{q(\beta)}{q(z)}=-\frac{1}{N^2}\frac{\sqrt{\beta}\sqrt{z}}{\sqrt{\beta}-\sqrt{z}}\frac{p(\beta)}{p(z)} \ . \label{eq:inverT} 
\end{align}
As promised, the square-root products $q$ defined in Eq.~(\ref{eq:deffz}), Eq.~(\ref{eq:defgz}) appear, and we have used the relation Eq.~(\ref{eq:refg}) to express them in terms of $p(\beta)$ and $p(z)$ defined in Eq.~(\ref{eq:deffb}), Eq.~(\ref{eq:defgb}).

To proceed further, we need the following summation formula
\begin{align}
\sum_{i,j}\frac{1}{\eta+y_j}\left(\frac{1}{x_i+y_j}\right)^{-1}_{ji}\frac{1}{x_i+\zeta}=\frac{1}{\eta+\zeta}-\frac{1}{\eta+\zeta}\frac{\prod_i (\eta-x_i) \prod_j(\zeta-y_j)}{\prod_j(\eta+y_j)\prod_i(\zeta+x_i)} \ , \label{eq:sumrule}
\end{align}
where $\eta \notin \{-y_j\}_j$ and $\zeta\notin\{- x_i\}_i$ are two extra numbers. This formula can be obtained from the standard inversion formula for Cauchy matrices by enlarging the original Cauchy matrix $(x_i+y_j)^{-1}$ with one extra column and row into a larger Cauchy matrix, with two new elements $\eta,\zeta$, and express its inverse at the $\zeta,\eta$ location in two different ways. By taking limits on $\eta$ and $\zeta$, this formula allows to compute all the summations required by this work into products. We first consider the product $HT^{-1}$. This quantity is also required in 
\begin{align}
&(v^TT^{-1})_\beta=\frac{i-1}{2}\left(\delta_{\beta 1}-(H T^{-1})_{1\beta}\right)  \ ,  \label{eq:vTb}\\ 
 &\langle v|T^{-1}|0\rangle=v_z(T^{-1})_{z1}=\frac{i-1}{2}\left(1
-(HT^{-1})_{11}\right) \ . \label{eq:vT0}
\end{align}
In Eq.~(\ref{eq:sumrule}), using $\eta=-\sqrt{\beta}$ and send $\zeta \rightarrow-\sqrt{\beta'}$, by comparing the residues one obtains the following 
\begin{align}
(HT^{-1})_{\beta\beta'}=-\frac{1}{\sqrt{\beta}+\sqrt{\beta'}}p(\beta)q(\beta')=-\frac{1}{N}\frac{\sqrt{\beta'}}{\sqrt{\beta}+\sqrt{\beta'}}p(\beta)p(\beta') \ , \label{eq:HTmius}
\end{align}
where the square root products appear again. Given the above,  the invertibility of $W$ can be addressed. From Eq.~(\ref{eq:relWT}), it is easy to show that 
\begin{align}
\det W=(1+\langle v|T^{-1}|0\rangle)\times \det T \ , 
\end{align}
since $T$ is invertible. From the Eq.~(\ref{eq:vT0}), Eq.~(\ref{eq:HTmius}) and Eq.~(\ref{eq:pq1}), one has 
\begin{align}
1+\langle v|T^{-1}|0\rangle=1+\frac{i-1}{2}+\frac{i-1}{4}p(1)q(1)=1+i \ne0 \ . \label{eq:creteri}
\end{align}
This crucial identity shows that $W$ is indeed invertible, justifying the overall strategy.

Knowing that $W$ is invertible, we proceed to compute all the overlapping matrices. First, for the $v$, one needs
\begin{align}
    W^{-1}=T^{-1}-\frac{1}{1+\langle v|T^{-1}|0\rangle} T^{-1} |0\rangle\langle v|T^{-1} \ .
\end{align}
Given $W^{-1}$, we can also compute 
\begin{align}
W'W^{-1}=HT^{-1}-\frac{1}{1+\langle v|T^{-1}|0\rangle}(|0\rangle+HT^{-1}|0\rangle)\langle v|T^{-1} \ , \label{eq:WprimW}
\end{align}
and 
\begin{align}
W^{-1}W'=T^{-1}H-\frac{1}{1+\langle v|T^{-1}|0\rangle}T^{-1}|0\rangle\left(\langle v|+\langle v|T^{-1}H\right) \ . 
\end{align}
In the above, $T^{-1}$, $HT^{-1}$ , $v^TT^{-1}$ and $\langle v|T^{-1}|0\rangle$ are already known. To finish derivation, it remains to compute the $T^{-1}H$ and $\langle v|T^{-1}H$. The former can be computed similar as $HT^{-1} $, while for the latter, one needs to use the sum-rule Eq.~(\ref{eq:sumrule}) with both $\zeta$ and $\eta$. The results read
\begin{align}
&(T^{-1}H)_{zz'}=\frac{1}{N}\frac{\sqrt{z}}{\sqrt{z}+\sqrt{z'}}\frac{1}{p(z)p(z')} \ , \\
&v_z+(v^TT^{-1}H)_z=-\frac{i-1}{2}\frac{p(1)}{(\sqrt{z}-1)p(z)} \ . 
\end{align}
We now have all the ingredients to obtain the final results for the $g$, $\hat g$ and $v$ defined in Eq.~(\ref{eq:defg}). 

We first present the results of $g$. Combining the Eq.~(\ref{eq:inverT}), Eq.~(\ref{eq:vTb}), Eq.~(\ref{eq:HTmius}), Eq.~(\ref{eq:creteri}) and Eq.~(\ref{eq:WprimW}) one has 
\begin{align}
&g_{\beta \beta'}=\left(W'W^{-1}\right)_{\beta \beta'}\nonumber \\
&=-\frac{p(\beta)q(\beta')}{\sqrt{\beta}+\sqrt{\beta'}}-\frac{i-1}{2(1+i)}\bigg(\delta_{\beta,1}-\frac{p(\beta)q(1)}{\sqrt{\beta}+1}\bigg)\bigg(\delta_{\beta',1}+\frac{p(1)q(\beta')}{\sqrt{\beta'}+1}\bigg)  \ . \label{eq:ggeneral}
\end{align}
We can write the individual components. For both $\beta'\ne 1$ and $\beta \ne 1$, one has 
\begin{align}
g_{\beta\beta'}=-\frac{\sqrt{\beta'}p(\beta)p(\beta')(1+\sqrt{\beta} \sqrt{\beta'})}{N(1+\sqrt{\beta})(1+\sqrt{\beta'})(\sqrt{\beta}+\sqrt{\beta'})} \ ;  \ \beta ,   \beta'\ne 1 \ ,
\end{align}
where we have used the relations $p(1)=(1-i)\sqrt{N}$, $q(1)=(1-i)\frac{1}{\sqrt{N}}$.
With one zero mode, one has 
\begin{align}
g_{1,\beta}=-\frac{\sqrt{\beta}p(\beta)}{\sqrt{N}(1+\sqrt{\beta})} \ ,  \ g_{\beta,1}=i\frac{p(\beta)}{\sqrt{N}(1+\sqrt{\beta})}   \ ; \  \beta\ne 1 \ . 
\end{align}
For both $\beta=\beta'=1$, from the Eq.~(\ref{eq:ggeneral}) and the relation $p(1)q(1)=-2i$, one can compute 
\begin{align}
g_{11}=-\frac{p(1)q(1)}{2}+\frac{1-i}{2(1+i)} \left(1-\frac{p(1)q(1)}{2}\right)\left(1+\frac{p(1)q(1)}{2}\right)\equiv 0 \ , 
\end{align}
as expected. Given $g_{\beta \beta'}$, to convert to the $u(\beta,\beta')=g_{\beta \bar \beta'}$, it is helpful to use the following relation for ${\cal Z} \in S_{\pm },\  {\cal Z} \ne 1$
\begin{align}
p\left(\bar {\cal Z}\right)=-i\frac{\sqrt{{\cal Z}}-1}{\sqrt{{\cal Z}}+1}p({\cal Z}) \ ,  \label{eq:idp}
\end{align}
which follows from Eq.~(\ref{eq:integralp}) and the fact that $\hat \Gamma_N({\cal Z})=\hat \Gamma(\bar {\cal Z})$ when ${\cal Z} \in S_{\pm }$. Using this relation, we obtain the desired overlapping matrix in Eq.~(\ref{eq:defu1}), where one of the $\beta, \beta'$ are not equal to $1$
\begin{align}
&\langle c_\beta c_{\beta'}\rangle =u(\beta,\beta')=g_{\beta \bar \beta'}=\frac{i}{N}\frac{\hat p(\beta)\hat p(\beta')}{(1+\sqrt{\beta})(1+\sqrt{\beta'})}\frac{\sqrt{\beta}-\sqrt{\beta'}} {\sqrt{\beta}\sqrt{\beta'}-1}  \ , \label{eq: resultu1} \\
&\hat p(\beta)=(1+i\delta_{\beta ,1})p(\beta) \ ,  \ \hat p(1)=2\sqrt{N} \ . 
\end{align}
As expected, it is manifestly skew-symmetric. Notice that $u(1,1)=g_{11}=0$. One can also obtain the other two overlapping matrices defined in Eq.~(\ref{eq:defu2}) and Eq.~(\ref{eq:defu3}) in a way similar to $u(\beta,\beta')$, and we will omit the computation details. For the $\hat g$, one has 
\begin{align}
&\hat g_{z,z'}=-\frac{1}{N}\frac{\sqrt{z}}{\sqrt{z}+\sqrt{z'}}\frac{1}{p(z)p(z')}-\frac{1}{N}\frac{\sqrt{z}}{(\sqrt{z}-1)(\sqrt{z'}-1)p(z)p(z')} \nonumber \\ 
&=-\frac{\sqrt{z}(\sqrt{z}\sqrt{z'}+1)}{N p(z)p(z')(\sqrt{z}-1)(\sqrt{z'}-1)(\sqrt{z}+\sqrt{z'})} \ ,  \label{eq:hatg}
\end{align}
leading to the skew-symmetric expression
\begin{align}
\langle c_z^\dagger c_{z'}^\dagger\rangle=u(z,z')=-\frac{i}{N}\frac{\sqrt{z}\sqrt{z'}}{p(z)p(z')(\sqrt{z}-1)(\sqrt{z'}-1)}\frac{\sqrt{z}-\sqrt{z'}}{\sqrt{z}\sqrt{z'}-1} \ . \label{eq:resultsu2}
\end{align}
Finally, for the mixed contraction, one has 
\begin{align}
\langle c_\beta c_z^\dagger\rangle=u(\beta,z)=-\frac{1}{N}\frac{\sqrt{z} \hat p(\beta)}{p(z)(\sqrt{z}-1)(\sqrt{\beta}+1)}\frac{\sqrt{z}\sqrt{\beta}-1}{\sqrt{\beta}-\sqrt{z}} \ . \label{eq:resultu3}
\end{align}
The overlapping matrices Eq.~(\ref{eq: resultu1}), Eq.~(\ref{eq:resultsu2}), Eq.~(\ref{eq:resultu3}) are the major results of this subsection. 

We now compute the contraction between the $\Gamma_0$ and the fermion operators. We start from contractions with $c_\beta$, and write
\begin{align}
\langle c_\beta\Gamma_0 \rangle=\frac{i}{\sqrt{N}}\left(1-\sum_{\beta'\in S_+}g_{\beta\beta'}\right) \ .  
\end{align}
To perform the sum, one needs the following summation rule
\begin{align}
\sum_{i,j}\frac{1}{\eta+y_j}\left(\frac{1}{x_i+y_j}\right)^{-1}_{ji}=1-\frac{\prod_i (\eta-x_i)}{\prod_j(\eta+y_j)} \ , 
\end{align}
which can be obtained from Eq.~(\ref{eq:sumrule}) by taking the large $\zeta$ limit. Using this, one has 
\begin{align}
&\sum_{\beta'\in S_+}(HT^{-1})_{\beta\beta'}=1-p(\beta)  \ , \  \sum_{\beta' \in S_+}(v^TT^{-1})_{\beta'}=\frac{i-1}{2}p(1) \ . 
\end{align}
This way, one obtains the crucial contraction
\begin{align}
\langle c_\beta \Gamma_0 \rangle=\frac{i}{\sqrt{N}}\frac{\sqrt{\beta} \hat p(\beta)}{1+\sqrt{\beta}} \ . \label{eq:contrcG}
\end{align}
In particular, for $\beta=1$, one has the simple relation for the one-point function
\begin{align}
\langle c_1 \Gamma_0 \rangle=\frac{\langle \Omega_{R}|c_1\sigma^x_0|\Omega_{NS}\rangle}{\langle \Omega_R |\Omega_{NS}\rangle}=i\ . \label{eq: onpointzero}
\end{align}
Similarly, the contraction with the $b_z^\dagger$ on the right side can also be computed 
\begin{align}
\langle \Gamma_0 c_{z}^{\dagger} \rangle=\frac{i}{\sqrt{N}}\bigg(-1+\sum_{z' \in S_-}\hat g_{z'\bar z}\bigg) =\frac{1}{\sqrt{N}}\frac{1}{\sqrt{z}-1}\frac{1}{p (z)} \ . \label{eq:contrGb} 
\end{align}
We note that the results in the subsection have been verified by numerical evaluation of $W^{-1}W'$, $W'W^{-1}$ and $W^{-1}$ at various $N$.

As a by-product of the overlapping matrices, we can also compute the phase difference $e^{i(\theta_+-\theta_-)}$ in Eq.~(\ref{eq:traphase}). Using $\sigma^z_0=-i\Gamma_0\Gamma_1$ and the mode decomposition of $\Gamma_0$ in terms of $c_\beta,c_\beta^\dagger$, $\Gamma_1$ in terms of $c_z$ and $c_z^\dagger$, one has 
\begin{align}
e^{-i(\theta_+-\theta_-)}=-\frac{\langle \Omega_{R}|i\Gamma_0\Gamma_1|\Omega_{NS}\rangle}{\langle \Omega_R|\Omega_{NS}\rangle}=-i\sum_{\beta\in S_+,z\in S_-}\varphi(z)(W^{-1})_{z\beta}\frac{\varphi(\beta)}{\beta} \ . 
\end{align}
All the sums can still be performed using Eq.~(\ref{eq:sumrule}). To obtain $\frac{1}{\sqrt{\beta}}$ in the denominator, one can simply set $\zeta=0$. However, to obtain the $\sqrt{z}$ in the numerator, one needs to expand at large $\eta$ to the second order
\begin{align}
\frac{1}{\eta+y_j}=\frac{1}{\eta}-\frac{y_j}{\eta^2}+{\cal O}\left(\frac{1}{\eta^3}\right) \ . 
\end{align}
At the end, one finds
\begin{align}
e^{-i(\theta_+-\theta_-)}=1 \ . 
\end{align}
This guarantees that there will be no extra $(-1)^i$ in the matrix elements of $\sigma^x_i$.

Here we comment on the $i$ in the one-point function Eq.~(\ref{eq: onpointzero}). As the ground states $|\Omega_{R}\rangle$ and $|\Omega_{NS}\rangle$ can be chosen purely real, this sign must due to the normalization of $c_1$ and $c_1^\dagger$. Let's show this. In our convention, from Eq.~(\ref{eq:convertc}) one has 
\begin{align}
c_1=\frac{i}{2\sqrt{N}}\sum_{i=0}^{N-1}\left(\Gamma_{2i}+i\Gamma_{2i+1}\right) \ ,  \ 
c_1^\dagger=\frac{-i}{2\sqrt{N}}\sum_{i=0}^{N-1}\left(\Gamma_{2i}-i\Gamma_{2i+1}\right) \ .  
\end{align}
Since the ground state $|\Omega_{R}\rangle$ is annihilated by $c_1$, one has
\begin{align}
\langle \Omega_R|c_1 \Gamma_0|\Omega_{NS}\rangle=\frac{i}{\sqrt{N}}\langle\Omega_R|\sum_{i=0}^{N-1}\Gamma_{2i} \Gamma_0|\Omega_{NS}\rangle \ .
\end{align}
Then, since it is possible to have all the $\Gamma_{2i}$ and $i\Gamma_{2i+1}$ purely real, the one-point function in Eq.~(\ref{eq: onpointzero}) indeed must be purely imaginary in our convention. Clearly, this $i$ can be traced back to the choice $u=i$ in Eq.~(\ref{eq:choiceunu}). The reason of making this choice is that it allows the mode decompositions Eq.~(\ref{eq:mode2}), Eq.~(\ref{eq:mode4}) of $\Gamma_{2i+1}$ to be expressed in terms of the Teoplitz symbol $C(z)$ without extra $i$s.

\subsection{The Wick contraction}
In this subsection, given all the three overlapping matrices in Eq.~(\ref{eq: resultu1}), Eq.~(\ref{eq:resultsu2}), Eq.~(\ref{eq:resultu3}), and the contraction with $\Gamma_0$ in Eq.~(\ref{eq:contrcG}), Eq.~(\ref{eq:contrGb}), we perform the Wick contraction to obtain the main result Eq.~(\ref{eq:formresult}) for the form factors. As a reminder, the matrix elements in Eq.~(\ref{eq:defM}) are defined as 
\begin{align}
{\cal M}_{2k+1,2l}(\{\beta_i\},\{z_j\})={\cal M}_{2k+1,2l}=(-1)^l\langle \prod_{i=1}^{2k+1}c_{\beta_i}\hat \sigma \prod_{j=1}^{2l}c_{z_j}^{\dagger}\rangle \ . 
\end{align}
The extra factor of $(-1)^l$ can be achieved through the redefinition
\begin{align}
c_z^{\dagger} \rightarrow -i c_z^{\dagger} \ , \  c_z \rightarrow ic_z \ , 
\label{eq:opered}
\end{align}
The advantage of this redefinition is that it makes the contractions among the two sectors more symmetric.

To perform the contraction, for the moment it is convenient to defined the following modified leg functions 
\begin{align}
&f_+(\beta)=\frac{\hat p(\beta)}{\sqrt{N}(1+\sqrt{\beta})}=\frac{d(\beta)}{\sqrt{\beta}} \ , \\ 
&\ f_-(z)=\frac{\sqrt{z}}{\sqrt{N}(\sqrt{z}-1)p(z)}=\sqrt{z}d(z) \ . 
\end{align}
It is easy to check after the operator redefinition Eq.~(\ref{eq:opered}), in every contraction contributing to the ${\cal M}_{2k+1,2l}$, after factorizing out a product of all the $f_+(\beta)$ and $f_-(z)$s, one obtains a $(2k+2l+2) \times (2k+2l+2)$ Pfaffian, whose entries are given by 
\begin{align}
&A(\beta,\beta')=i\frac{\sqrt{\beta}-\sqrt{\beta'}}{\sqrt{\beta}\sqrt{\beta'}-1} \ ,  \  A(z,z')=i\frac{\frac{1}{\sqrt{z}}-\frac{1}{\sqrt{z'}}}{\frac{1}{\sqrt{z}}\frac{1}{\sqrt{z'}}-1} \ , \\
&A(\beta,\sigma)=-i\frac{\sqrt{\beta}}{0-1} \ , \ A(\sigma,z)=-i\frac{0-\frac{1}{\sqrt{z}}}{-1} \ ,\\
&A(\beta,z)=i\frac{\sqrt{\beta}-\frac{1}{\sqrt{z}}}{\sqrt{\beta}\frac{1}{\sqrt{z}}-1} \ . 
\end{align}
Here, $A(\beta,\sigma)$ and $A(\sigma,z)$ denote the contractions with the spin field. As such, if we introduce the following numbers 
\begin{align}
&s_i=\sqrt{\beta_i} \ , 1\le i\le 2k+1 \ ; s_{2k+2}=0  \ ; \  s_{2k+2+j}=\frac{1}{\sqrt{z_j}} ,  \ 1\le j\le 2l \ ,
\end{align}
the matrix element can be expressed as 
\begin{align}
{\cal M}_{2k+1,2l}=-i^{k+l+1}\prod_{1\le i\le 2k+1}f_+(\beta_i)\prod_{1\le j\le 2l}f_-(z_j) {\rm Pfaff} \left(\frac{s_i-s_j}{s_is_j-1}\right)_{1\le i,j\le 2k+2l+2} \ .
\end{align}
Here, the factor $i^{k+l+1}$ is due to the $i$s in each term of the contraction, and the overall minus sign is due to the sign flip for contractions with the $s_{2k+2}$. The crucial determinant formula is that for $n\in Z_{\ge1}$, one has 
\begin{align}
{\rm Pfaff}\left(\frac{s_i-s_j}{s_is_j-1}\right)_{1\le i, j \le 2n }=\prod_{1\le i<j\le 2n} \frac{s_i-s_j}{s_is_j-1} \ . \label{eq:pfaff}
\end{align}
To show this, we can change variables as 
\begin{align}
s_i=\frac{1+t_i}{1-t_i}, \  
\frac{s_i-s_j}{s_is_j-1}\rightarrow\frac{t_i-t_j}{t_i+t_j} \ , 
\end{align}
and then the Eq.~(\ref{eq:pfaff}) becomes a Cauchy-type identity. Using Eq.~(\ref{eq:pfaff}), the contractions with the spin operator leads to an extra $-1$ sign, as well as a product of all the $\frac{1}{\sqrt{z}}$s from the $NS$ sector, and another of all the $\sqrt{\beta}$s from the R sector. These factors compensate the overall minus sign, and the additional factors in $f_{\pm}$. As such, one obtains 
\begin{align}
&(-1)^l\langle \prod_{i=1}^{2k+1}c_{\beta_i}\hat \sigma \prod_{j=1}^{2l}c_{z_j}^{\dagger}\rangle={\cal M}_{2k+1,2l}(\{\beta_i\},\{z_j\})\nonumber \\ 
&=i^{k+l+1}\prod_{i<i'}\frac{\sqrt{\beta_i}-\sqrt{\beta_j}}
{\sqrt{\beta_i}\sqrt{\beta_j}-1}\prod_{i,j}\frac{\sqrt{\beta_i}\sqrt{z_j}-1}{\sqrt{\beta_i}-\sqrt{z_j}}\prod_{j<j'}\frac{\sqrt{z_j}-\sqrt{z_{j'}}}{\sqrt{z_j}\sqrt{z_{j'}}-1} \times \prod_i d(\beta_i)\prod_j d(z_j) \ . 
\end{align}
This exactly the desired result Eq.~(\ref{eq:formresult}).

\subsection{The vacuum overlap}

As an application of the form factors, in this subsection we compute the overlap $\langle \Omega_R|\Omega_{NS}\rangle$
between the two ground states and prove the Eq.~(\ref{eq:overlap}). First notice that for any two $z_i$ and $z_j$ in $S_-$, the quantity $\frac{\sqrt{z_i}-\sqrt{z_j}}{\sqrt{z_i}\sqrt{z_j}-1}$ is real. As such, the following spin-spin two point correlator ($\sigma(\tau)=e^{H\tau}\sigma^x_0e^{-H\tau}$) in the imaginary time direction can be expressed as
\begin{align}
&e^{-\Delta_E(N)\tau}\langle \Omega_{R}|c_1\sigma(\tau)\sigma(0)c_1^\dagger|\Omega_{R}\rangle\nonumber \\ 
&=|\langle \Omega_R|\Omega_{NS}\rangle|^2\times \sum_{k=0}^{\frac{N}{2}}\frac{1}{(2k)!}\sum_{z_i\in S_-}\prod_i |d(z_i)|^2e^{-\epsilon(z_i)\tau} \prod_{i<j} \left(\frac{\sqrt{z}_i-\sqrt{z_j}}{\sqrt{z_i}\sqrt{z_j}-1}\right)^2 \ . \label{eq:2pt1}
\end{align}
We can introduce the following $N\times N$ skew-symmetric matrix
\begin{align}
K_\tau(z,z')=\frac{|d(z)d(z')|(\sqrt{z}-\sqrt{z'})}{(\sqrt{z}\sqrt{z'}-1)}e^{-\frac{\epsilon(z)+\epsilon(z')}{2}\tau}   \ , 
\end{align}
then using Eq.~(\ref{eq:pfaff}), one has 
\begin{align}
e^{-\Delta_E(N)\tau}\langle \Omega_{R}|c_1 \sigma(\tau)\sigma(0)c_1^\dagger|\Omega_{R}\rangle=|\langle \Omega_R|\Omega_{NS}\rangle|^2\det (1+K_\tau) \ ,
\end{align}
where the $1$ in such Fredholm determinants always means the identity operator. In particular, for $\tau=0$, since $(\sigma^x_0)^2=1$, one has 
\begin{align}
1=|\langle \Omega_R|\Omega_{NS}\rangle|^2\det (1+K_0) \ , 
\end{align}
which allows the overlap between the two ground states to be computed as a Fredholm determinant. However, one also has 
\begin{align}
&\det(1+K_0)=\det(1+W^{-1}W') \ , \label{eq:matrixiden}
\end{align}
where $W$ and $W'$ are defined in Eq.~(\ref{eq:defW}), and the matrix elements for $\hat g=-W^{-1}W'$ are given explicitly in Eq.~(\ref{eq:hatg}).
To show that these determinants are the same, we choose the following particular ordering for the $z\in S_-$
\begin{align}
z_0=e^{i\frac{\pi}{N}}, z_1=e^{-i\frac{\pi}{N}}.... \ , z_{2k}=e^{i\frac{2k+1}{N}\pi}, z_{2k+2}=e^{-i\frac{2k+1}{N}\pi} \ , \  0\le k\le \frac{N-2}{2} \ . 
\end{align}
In this ordering, we simply notice the following identity
\begin{align}
Q^{-1}(W^{-1}W')Q =iK_0\Sigma\ , \   Q_{z,z'}=\delta_{z,z'}\sqrt{\sqrt{z}} \ , \ \Sigma= \otimes_{i=0}^{\frac{N-2}{2}}\sigma_1  \ . \label{eq:K0Wrele}
\end{align}
This can be proven using the identity Eq.~(\ref{eq:idp}) for $p(z)$ and the fact that $p(z)=e^{-\frac{i\pi}{4}}|p(z)|$ when $z\in S_-$. More precisely, multiplying $\Sigma$ from the right switches $z$ and $\bar z$ in the columns in our ordering, while for $z,z'\in S_-$, one always has 
\begin{align}
\frac{\sqrt{\sqrt{z}}\sqrt{-\frac{1}{\sqrt{z'}}}\sqrt{z'}}{(\sqrt{z}-1)(\sqrt{z'}-1)}=\frac{-i}{|(\sqrt{z}-1)(\sqrt{z'}-1)|} \ , 
\end{align}
if all the square-roots are defined using the $(0,2\pi)$ branch. The above are sufficient to show the identity Eq.~(\ref{eq:K0Wrele}).  As such, one has 
\begin{align}
\det(1+W^{-1}W')=\det(1+iK_0\Sigma) \ . \label{eq:matr1}
\end{align}
However, one also has
\begin{align}
\Sigma K_0\Sigma=-K_0 \ . \label{eq:matr2}
\end{align}
This is because the Eq.~(\ref{eq:idp}) implies that $|p(z)(\sqrt{z}-1)|=|p(\bar z)(\sqrt{\bar z}-1)|$. Combining Eq.~(\ref{eq:matr1}) and Eq.~(\ref{eq:matr2}) establishes the identity Eq.~(\ref{eq:matrixiden}).

We now compute the determinant of $1+W^{-1}W'$ as
\begin{align}
\det(1+W^{-1}W')=\det W^{-1} \det (W+W')=\det \left(\frac{2\varphi_z}{\beta-z}\right)\det \left(\frac{\varphi_\beta+\varphi_z}{\beta-z}\right)^{-1} \ . 
\end{align}
For $W$, using the definition Eq.~(\ref{eq:defW}), the relations Eq.~(\ref{eq:relWT}) and the crucial identity Eq.~(\ref{eq:creteri}), we can write
\begin{align}
\det W=\det (T+|0\rangle\langle v|)=\det T (1+\langle v|T^{-1}|0\rangle)=(1+i)\det T \ . 
\end{align}
As such, one has 
\begin{align}
1=|\langle \Omega_R|\Omega_{NS}\rangle|^2\times \frac{2^N(-1)^{\frac{N}{2}}}{1+i}\det \frac{1}{\beta-z} \det  \left(\frac{1}{\sqrt{\beta}-\sqrt{z}}\right)^{-1} \ . 
\end{align}
Thus, it is now a ratio of two Cauchy determinants. We first consider the simple one, the inverse of which read
\begin{align}
T^0_{\beta z}= \frac{1}{\beta-z} ,  \  
(T^0)^{-1}_{z\beta}=-\frac{1}{\beta-z} \frac{\prod_{z'\in S_-} (\beta-z')}{\prod_{\beta'\ne \beta}(\beta-\beta')}\frac{\prod_{\beta'\in S_+ }(z-\beta')}{\prod_{z'\ne z}(z-z')} \ . 
\end{align}
We already encountered these unrestricted products before when showing the Eq.~(\ref{eq:refg})
\begin{align}
\sum_{z'\in S_-} \ln \left(1-\frac{z'}{\beta}\right)= \ln 2 \ , \  
\sum_{\beta' \in S_+,\beta'\ne \beta} \ln \left(1-\frac{\beta'}{\beta}\right)=\ln N \ , \\
\sum_{\beta'\in S_+}\ln \left(1-\frac{\beta'}{z}\right)=\ln 2 \ ,   \ 
\sum_{z'\in S_-,z'\ne z}\ln \left(1-\frac{z'}{z}\right)=\ln N \ . 
\end{align}
As such, one has 
\begin{align}
(T^0)_{z\beta}^{-1}=-\frac{4}{N^2}\frac{\beta z}{\beta -z} \ . 
\end{align}
From the inverse, the determinant can be easily extracted. Since the $|\langle\Omega_R |\Omega_{NS}\rangle|^2$ is positive, here we will not care too much about the signs
\begin{align}
\frac{1}{\det T^0}=\det T^0\times\left(\frac{4}{N^2}\right)^N\times (-1) \ ,  \  \det T^0=\pm i\left(\frac{N}{2}\right)^{N} \ . 
\end{align}
This determinant is very large. Now, it comes to the matrix $T$ defined in Eq.~(\ref{eq:defTH}). Using the inversion formula Eq.~(\ref{eq:inverT}), one can write
\begin{align}
\frac{1}{\det T}=\frac{\det T}{N^{2N}} \prod_{\beta} \sqrt{\beta} \prod_{z} \sqrt{z} \prod\frac{p(\beta)}{p(z)}=-i\frac{\det T}{N^{2N}}\prod\frac{p(\beta)}{p(z)} \ . 
\end{align}
Here, using the Eq.~(\ref{eq:integralp}) and Eq.~(\ref{eq:exactfg1}) we can further simplify
\begin{align}
\prod\frac{p(\beta)}{p(z)} = \sqrt{2N}\exp \left(\sum_{\beta\ne 1}\hat \Gamma_N(\beta)-\sum_{z\in S_-}\hat \Gamma_N(z)\right)=P>0 \ , 
\end{align}
thus one has 
\begin{align}
\det T^2=iN^{2N}P^{-1} \ ,  \  \det T=\pm \frac{1}{\sqrt{2P}}(1+i)N^N \ ,
\end{align}
leading to
\begin{align}
\det(1+W^{-1}W')=\frac{\sqrt{2P}}{2}=\frac{\sqrt{2}}{2}(2N)^{\frac{1}{4}}\times \exp \left(\frac{1}{2}\sum_{\beta\ne 1}\hat \Gamma_N(\beta)-\frac{1}{2}\sum_{z\in S_-}\hat \Gamma_N(z)\right) \  .
\end{align}
This gives the following ground state overlap
\begin{align}
\bigg|\langle \Omega_R|\Omega_{NS}\rangle\bigg|^2=\left(\frac{2}{N}\right)^{\frac{1}{4}}\exp \left(\frac{1}{2}\sum_{z\in S_-}\hat \Gamma_N(z)-\frac{1}{2}\sum_{\beta \in S_+, \  \beta\ne 1}\hat \Gamma_N(\beta)\right) \ . 
\end{align}
This finalizes the derivation for the overlap in Eq.~(\ref{eq:overlap}).

We now derive the integral representation Eq.~(\ref{eq:intlap}) that allows to expand the overlap to an arbitrary order in $\frac{1}{N}$. We start from Eq.~(\ref{eq:integralp}). Here the key observation is that for $\beta \in S_+/\{1\}$, as well as for $z\in S_-$, one always has ${\rm Re}(-i\sqrt{z}), {\rm Re}(-i\sqrt{\beta)}>0$. As such, it is convenient to introduce a Barnes representation in the variable $-i\sqrt{\beta}$. Denote the Barnes parameter as $u$, by performing the Mellin transform under the integral representations Eq.~(\ref{eq:defGamma}), for ${\cal Z}\notin R_+$ one has
\begin{align}
\hat \Gamma_N({\cal Z})=\int_0^1dt \int \frac{du}{2\pi i} \frac{t^{-1-\frac{u}{2}}(1+t^u)}{4\cos \frac{\pi u}{2}}\ln \frac{1-t^N}{1+t^N} (-i\sqrt{\cal Z})^{-u} \ , -1<{\rm Re}(u)<1 \ . 
\end{align}
Now, as far as ${\cal Z}\notin R_+$, the exponential decay in the ${\rm Im}(u)$ direction will be preserved and the double integral is absolutely convergent. The Barnes representation also simplifies the $t$ dependencies, allowing the $t$ integral to be performed explicitly as
\begin{align}
\hat \Gamma_N({\cal Z})=-\int \frac{du}{4i } \frac{\tan \frac{\pi u}{4N}}{u \cos \frac{\pi u}{2}} (-i\sqrt{\cal Z})^{-u} \ .
\end{align}
It is easy to perform the finite sums which are geometric and obtain 
\begin{align}
&\frac{1}{2}\sum_{z\in S_-} \hat \Gamma_N(z)-\frac{1}{2}\sum_{\beta\in S_+, \beta \ne 1} \hat \Gamma_N(\beta) ={\cal I=}\int \frac{du}{16i  u} \frac{\sin \frac{(N-1)\pi u}{2N}-\sin \frac{\pi u }{2}}{\cos \frac{\pi u}{2}\cos ^2\frac{\pi u}{4N}} \ . 
\end{align}
Now, to finally obtain the Eq.~(\ref{eq:intlap}), we choose the integration path along the imaginary axis $u=ip$ 
\begin{align}
{\cal I}=-\int_{-\infty}^{\infty} dp\frac{\left(e^{\frac{\pi  p}{2 N}}-1\right) \left(e^{\frac{\pi  p}{2 N}}+e^{\pi  p}\right)}{4p \left(e^{\pi  p}+1\right)  \left(e^{\frac{\pi  p}{2 N}}+1\right)^2}=-\int_{0}^{\infty}dp\frac{\left(e^{\frac{\pi  p}{2 N}}-1\right) \left(e^{\frac{\pi  p}{2 N}}+e^{\pi  p}\right)}{2p \left(e^{\pi  p}+1\right)  \left(e^{\frac{\pi  p}{2 N}}+1\right)^2}  \ . 
\end{align}
The second equality is because the integrand is symmetric under $p \rightarrow -p$. Finally, we change the variable to $t=e^{-\frac{\pi p}{n}}$ and obtain
\begin{align}
{\cal I}=\int_0^1 dt \frac{(1-\sqrt{t}) (t^N+\sqrt{t})}{(1+\sqrt{t})^2  \left(t^N+1\right)}\frac{1}{2t\ln t} \ , 
\end{align}
which is exactly Eq.~(\ref{eq:intlap}). To perform the large $N$ expansion, one can write
\begin{align}
&\int_0^1 dt \frac{(1-\sqrt{t}) (t^N+\sqrt{t})}{(1+\sqrt{t})^2  \left(t^N+1\right)}\frac{1}{2t\ln t}=\int_0^1 dt \frac{(1-\sqrt{t}) }{(1+\sqrt{t})^2  }\frac{1}{2\sqrt{t}\ln t}
-\int_0^{\infty} dt \frac{\tanh^2 \frac{t}{4N} }{2t(e^t+1)} \ ,
\end{align}
where in the second term we changed the integration variable as $t \rightarrow e^{-\frac{t}{N}}$. The $N$-independent first term is the dominant contribution in the large $N$ limit. It can be evaluated exactly as  
\begin{align}
\int_0^1 dt \frac{(1-\sqrt{t}) }{(1+\sqrt{t})^2  }\frac{1}{2\sqrt{t}\ln t}=\ln \left(\pi^{\frac{1}{4}}G(\frac{1}{2})G(\frac{3}{2})\right)\equiv \ln \left(\frac{2^{\frac{1}{12}}(\pi e)^{\frac{1}{4}}}{A^3}\right) \ ,  \label{eq:Claiinte1}
\end{align}
where $A$ is the Glaisher constant. Eq.~(\ref{eq:Claiinte1}) will be shown in the Appendix.~\ref{sec:integral}.  On the other hand, the term with  $\tanh^2\frac{t}{4N}$ 
represents the sub-leading corrections and can be easily expanded to an arbitrary order in $\frac{1}{N}$
\begin{align}
-\int_0^{\infty} dt \frac{\tanh^2 \frac{t}{4N} }{2t(e^t+1)} =-\frac{1}{32N^2}\int_0^{\infty}\frac{tdt}{e^{t}+1}+\frac{1}{768N^4}\int_0^{\infty}\frac{t^3dt}{ e^{t}+1}+{\cal O}\left(\frac{1}{N^6}\right) \ . 
\end{align}
Integrals here are elementary and can be integrate to $\zeta$ values at positive even integers. As such, in the large $N$ limit, one has the following expansion
\begin{align}
\bigg|\langle \Omega_R|\Omega_{NS}\rangle\bigg|^2=\left(\frac{2\pi }{N}\right)^{\frac{1}{4}}G\left(\frac{1}{2}\right)G\left(\frac{3}{2}\right)\exp \bigg(\frac{-\pi ^2}{384 N^2}+\frac{7 \pi ^4}{92160 N^4}+{\cal O}\left(\frac{\pi^6}{N^6}\right)\bigg) \ , \label{eqL:oversubl}
\end{align}
confirming the asymptotics Eq.~(\ref{eq:ovlasym}). Notice that the $\frac{1}{N}$ corrections are quite small at the first few orders, but will finally grow factorially. Still, the large $N$ expansion is Borel summable.

At the end of this subsection, we would like to address two issues. The first issue is on the origin of the constant $(\pi)^{\frac{1}{4}}G(\frac{1}{2})G(\frac{3}{2})$. In the derivations above based on the exact integral representation Eq.~(\ref{eq:intlap}), this constant is due to the integral Eq.~(\ref{eq:Claiinte1}). However, from the asymptotics Eq.~(\ref{eq:Gammalim}) and the Binet's integral Eq.~(\ref{eq:binet}), in the scaling region, the $\hat \Gamma_N(\beta)$ and $\hat \Gamma_N(z)$ are all finite and allow simple expressions in terms of gamma functions. Is the Glaisher constant purely due to the scaling region's contribution to the sum? The answer is yes. In fact, one can show that 
\begin{align}
 \prod_{i=0}^{\infty} \sqrt{\frac{i+\frac{1}{2}}{i+1}}\frac{\Gamma(i+1)\Gamma(i+2)}{\Gamma^2(i+\frac{3}{2})}=\pi^{\frac{1}{4}}G\left(\frac{1}{2}\right)G\left(\frac{3}{2}\right) \ . \label{eq:product1}
\end{align}
To show the product formula in Eq.~(\ref{eq:product1}), simplify notice that products of factorials can be expressed through hyper-factorials $H(n)=\prod_{i=1}^n i^i$, whose large $n$ asymptotics lead to the Glaisher constant
\begin{align}
\sum_{i=1}^n\ln i!=(n+1)\ln n!-\ln \prod_{i=1}^ni^i=(n+1)\ln n!-\ln H(n) \ , \\
\ln H(n)\rightarrow \left(\frac{n^2+n}{2}+\frac{1}{12}\right)\ln n -\frac{n^2}{4}+\ln A \ . 
\end{align}
We should also mention that instead of using the Binet's integral Eq.~(\ref{eq:binet}), if one perform the sum in the Barnes representation Eq.~(\ref{eq:Barnb}), one can obtain the following Barnes representation for the Glaisher constant
\begin{align}
\int_{c-i\infty}^{c+i\infty} \frac{du}{2\pi i}(2-2^u)(2-2^{-u})\Gamma(u)\Gamma(-u)\zeta(u)\zeta(-u)=\ln \bigg(\pi^{\frac{1}{4}}G\left(\frac{1}{2}\right)G\left(\frac{3}{2}\right) \bigg) \ , \label{eq:Barnessymetric}
\end{align}
where $0<c<1$. Notice that the integrand is symmetric under $z\rightarrow -z$. To be more precise, one first write 
\begin{align}
&\sum_{i=0}^{\infty} \bigg(\mathfrak{b}(\pi(2i+1))-\mathfrak{b}(2\pi(i+1))\bigg) \nonumber \\ 
&=\sum_{i=0}^{\infty}\bigg(\ln \frac{\sqrt{i+\frac{1}{2}}}{\sqrt{i+1}}+\ln \frac{\Gamma\left(i+1\right)}{\Gamma(i+\frac{3}{2})}-\ln \frac{\Gamma\left(i+\frac{3}{2}\right)}{\Gamma(i+2)}\bigg)=\ln \left(\pi^{\frac{1}{4}}G\left(\frac{1}{2}\right)G\left(\frac{3}{2}\right)\right) \ . 
\end{align}
where the $\mathfrak{b}(\alpha)$ is defined in Eq.~(\ref{eq:binet}), and we have used Eq.~(\ref{eq:product1}). Then, due to the absolute convergence, we can sum under the Barnes integral Eq.~(\ref{eq:Barnb}). This leads to
\begin{align}
 &\sum_{i=0}^{\infty} \bigg(\mathfrak{b}(\pi(2i+1))-\mathfrak{b}(\pi(2i+2))\bigg)\nonumber \\ 
 &=-\int_{0<{\rm Re}(u)<1}\frac{du}{2\pi i}\left(1-\frac{1}{2^{u+1}}\right)\frac{\Gamma(u)\zeta(u+1)}{\cos \frac{\pi u}{2}}(2\pi)^{-u} \times (2^u-2)\zeta(u) \ . 
\end{align}
Finally, we use the functional equation of the $\zeta$ function to change $\zeta(u+1)$ to $\zeta(-u)$ and obtain Eq.~(\ref{eq:Barnessymetric}). In Appendix.~\ref{sec:expandgI}, as an application of this Barnes integral, we compute the small mass expansion of a related crossover function.

The second issue we would like to address is the computation of the sub-leading corrections. From the Eq.~(\ref{eq:Gammalim}), one can expand the $\hat \Gamma_N$ function in the scaling region to the sub-leading orders
\begin{align}
\hat \Gamma_N\left(e^{i\frac{2\pi k}{N}}\right)=\mathfrak{b}(2\pi|k|)-\frac{|k|\pi^2}{48N^2}+\frac{7 \pi ^4 |k| \left(1-8 k^2\right)}{23040 N^4}+{\cal O}\left(\frac{1}{N^6}\right) \ . 
\end{align}
On the other hand, after summing over $k\in Z$ and $k\in Z+\frac{1}{2}$, divergences are generated in the subleading terms, even after taking the difference between the two sectors. It is interesting to check, if the standard $\zeta$-regularization work in this case. For this, we compute
\begin{align}
&-\frac{\pi^2}{48N^2} \left(\sum_{m\in Z_{\ge 0}+\frac{1}{2}}\frac{1}{m^u}-\sum_{n\in Z_{\ge 1}}\frac{1}{n^u}\right)\bigg|_{u\rightarrow -1}=-\frac{\pi^2}{384N^2} \ , \\
&\frac{7 \pi ^4}{23040 N^4}\bigg(\sum_{m\in Z_{\ge0}+\frac{1}{2}}\frac{1-8 m^2}{m^u}- \sum_{n\in Z_{\ge 1}}\frac{1-8 n^2}{n^u}\bigg)\bigg|_{u\rightarrow-1}=\frac{7\pi^4}{92160N^4} \ . 
\end{align}
The first two subleading corrections in Eq.~(\ref{eqL:oversubl}) are exactly reproduced.  It is reasonable to expect that the $\zeta$ regularization works for higher order corrections as well.

\subsection{Scaling limit of the two-point correlator} \label{sec:scale}
As the $N\rightarrow \infty$, we have shown that the overlap between the two ground states approaches $0$ at a very slow speed $N^{-\frac{1}{4}}$ controlled by the scaling dimension of the spin operator. The rescaled spin operator $\hat \sigma$ in Eq.~(\ref{eq:defspinr}), on the other hand, allows the finite volume scaling limit at the level of the form factors, given by the Eq.~(\ref{eq:formCFT1}), Eq.~(\ref{eq:formCFT2}) and Eq.~(\ref{eq:phases}). Here we would like to show that the finite-volume scaling limit at the level of the spin-spin two-point correlator also exists and leads to the standard CFT correlator on a cylinder. More precisely, we first separate the two spin operators by an Euclidean time like in Eq.~(\ref{eq:2pt1}), but within two NS ground states. Then, the finite volume scaling limit means
\begin{align}
\tau \rightarrow +\infty \ , N\rightarrow +\infty , \  \frac{r}{L}=\frac{\tau}{N}={\cal O }(1) \ . 
\end{align}
We first show that the scaling limit of the form factor expansion can be taken under the summation for the individual form factors. For this, first notice that the time evolution factors can be bounded uniformly by  
\begin{align}
e^{-\tau \epsilon(\beta)} =e^{-\frac{r}{L}N\epsilon(\beta)} \le e^{-\frac{\pi r |k| }{L}},  \ \beta=e^{\frac{2\pi i k}{N}},  \  |k|\le \frac{N}{2} \ .
\end{align}
Second, notice that if $\beta, \beta' \in S_+/\{1\}$, then 
\begin{align}
\bigg|\frac{\sqrt{\beta}-\sqrt{\beta'}}{\sqrt{\beta}\sqrt{\beta'}-1}\bigg| \le 1 \ , 
\end{align}
which allows to bound the two-particle products in the form factor expansions by one. Finally, for the dressing functions, one needs just the simple bound
\begin{align}
|d(\beta)| \le c, 
\end{align}
where $c$ is an $N$-independent number. This way, after fixing $r>0$, the finite-$N$ form factor expansion is uniformly bounded from the above by an absolutely summable infinity sequence
\begin{align}
&e^{\Delta_E(N)\tau}\langle\Omega_{NS}| \hat\sigma(\tau)\hat\sigma(0)|\Omega_{NS}\rangle\nonumber \\ 
= \ &\sum_{k=0}^{\frac{N-2}{2}}\frac{1}{(2k+1)!}\sum_{\beta_i\in S_+/\{1\}}\prod_{i=1}^{2k+1}e^{-\epsilon(\beta_i)\tau} |d(\beta_i)|^2 \prod_{i<j} \left(\frac{\sqrt{\beta}_i-\sqrt{\beta_j}}{\sqrt{\beta_i}\sqrt{\beta_j}-1}\right)^2 
\nonumber \\ 
 +&1+ \sum_{k=1}^{\frac{N-2}{2}}\frac{1}{(2k)!}\sum_{\beta_i\in S_+/\{1\}}\prod_{i=1}^{2k}e^{-\epsilon(\beta_i)\tau} |d(\beta_i)|^2 \prod_{i<j} \left(\frac{\sqrt{\beta}_i-\sqrt{\beta_j}}{\sqrt{\beta_i}\sqrt{\beta_j}-1}\right)^2 
\nonumber \\ &\le \sum_{k=0}^{\infty} \frac{1}{k!}\left(2c^2\sum^{\infty}_{n=1} e^{-\frac{\pi r n}{L}}\right)^{k}<\infty \ . \label{eq:bounds}
\end{align}
Notice that the second line is the contribution from intermediate states without $\beta=1$ (zero momentum) excitations, while the third line is from states with a $\beta=1$ excitation. In particular, the $+1$ in the third line is due to the intermediate state $c_1^\dagger|\Omega_R\rangle$, and we have used the fact $d(1)=1$ in the third line. The bound Eq.~(\ref{eq:bounds}) allows to take the scaling limit under the sum, due to the dominated convergence theorem. 

After showing the convergence, we show that the rescaled two point correlator can be expressed as the square of a Fredholm determinant. In fact, in the product formulas  Eq.~(\ref{eq:formCFT1}), Eq.~(\ref{eq:formCFT2}), there are no correlations between the left and right momentum.  As such, in the vacuum two point correlator, within each of the four possible even odd parity assignments of the excitation numbers, the particle numbers in different L, R sectors can be summed independently, leading to 
\begin{align}
\langle \Omega_{NS}|\hat\sigma(r)\hat \sigma(0)|\Omega_{NS}\rangle=e^{-\frac{\pi r}{4L}}\left(2F_+F_-+F_+^2+F_-^2\right)=e^{-\frac{\pi r}{4L}}(F_++F_-)^2 \ .
\end{align}
Here, the $e^{-\frac{\pi r}{4L}}$ is due to the ground state energy difference Eq.~(\ref{eq:deltae}), while the $F_+, F_-$ are defined as 
\begin{align}
&F_+(w)=\sum_{k=0}^{\infty}\frac{1}{(2k)!}\sum_{\{n_i\ge 1\}}\prod_{i=1}^{2k}\tau^2(n_i)w^{n_i}\prod_{i<j} \left(\frac{n_i-n_j}{n_i+n_j}\right)^2 \ , \\
&F_-(w)=\sum_{k=0}^{\infty}\frac{1}{(2k+1)!}\sum_{\{n_i\ge 1\}}\prod_{i=1}^{2k+1}\tau^2(n_i)w^{n_i}\prod_{i<j} \left(\frac{n_i-n_j}{n_i+n_j}\right)^2  \ , 
\end{align}
where $\tau (n)$ is defined in Eq.~(\ref{eq:deftau}), and we have introduced the notation
\begin{align}
w=e^{-\frac{2\pi r}{L}} \ .
\end{align}
In fact, the $2F_+F_-$ is the scaling limit for the second line of Eq.~(\ref{eq:bounds}), while $F_+^2+F_-^2$ is for the third line. Using the Cauchy determinant formula, it is nor hard to show that the $F_++F_-$ can be combined to a single Fredholm determinant
\begin{align}
&F_+(w)+F_-(w)=\det (1+{\cal  K}_{ij})_{1\le i,j\le \infty} \ , \ {\cal K}_{ij}=\frac{1}{\pi}\frac{i}{i+j}\frac{\Gamma(i+\frac{1}{2})\Gamma(j+\frac{1}{2})}{\Gamma(i+1)\Gamma(j+1)} w^j  \ .
\end{align}
As such, one obtains the the following Fredholm determinant representation of the rescaled correlator
\begin{align}
\langle \Omega_{NS}|\hat \sigma(r)\hat \sigma(0)|\Omega_{NS}\rangle=e^{-\frac{\pi r}{4L}}\det (1+{\cal  K}_{ij})^2 \ . \label{eq:Fredholmscale}
\end{align}
This is another major result of the work. Notice that after the similarity transformation $iw^j \rightarrow \sqrt{ij} w^{\frac{i+j}{2}}$, the operator ${\cal K}$ is clearly of trace-class when $|w|<1$, and the Fredholm determinant is well defined. 

At small $w$, the Eq.~(\ref{eq:Fredholmscale}) allows to expand the scaling function to any given order in $w$. We have checked that up to $w^{30}$, the expansion coefficients exactly agree with that of the $(1-w)^{-\frac{1}{8}}$: 
\begin{align}
\det (1+{\cal  K}_{ij})=1+\frac{w}{8}+\frac{9w^2}{128}+\frac{51 w^3}{1024}+\frac{1275 w^4}{32768}+..... \ . 
\end{align}
It is very likely that this is an exact relation. Assuming this, one has then 
\begin{align}
\langle \Omega_{NS}|\hat \sigma(r)\hat \sigma(0)|\Omega_{NS}\rangle=e^{-\frac{\pi r}{4L}} \left(1-e^{-\frac{2\pi r}{L}}\right)^{-\frac{1}{4}}  \ . 
\end{align}
This is exactly the standard CFT two-point correlator on an infinite cylinder~\cite{DiFrancesco:1987cx,DiFrancesco:1987ez,DiFrancesco:1997nk} and generalizes the sum-rules in~\cite{DiFrancesco:1987ez}. As such, after multiplying back the factor $|\langle \Omega_R|\Omega_{NS} \rangle|^2$, at small $r\ll L$, one has then
\begin{align}
\left(\frac{2\pi }{N}\right)^{\frac{1}{4}}\times G\left(\frac{1}{2}\right)G\left(\frac{3}{2}\right) \times \left(\frac{2\pi r}{L}\right)^{-\frac{1}{4}} \rightarrow \frac{1}{\tau^\frac{1}{4}}G\left(\frac{1}{2}\right)G\left(\frac{3}{2}\right) \ , \  \frac{\tau}{N}=\frac{r}{L} \ . 
\end{align}
This is exactly the large distance asymptotics of the infinite-volume critical chain. As such, the simple fixed-point scaling scenario defined in~\cite{McCoy:2000gw} continues to hold in this example, if one believes that $\det (1+{\cal K})=(1-w)^{-\frac{1}{8}}$. 

It should be note that it is also possible to locate the two spin operators at different spatial positions, for example, one at $(\tau,i)$ and another at $(0,0)$, where $0\le i\le N-1$ is the spatial position in the lattice unit. Then the scaling limit in the spatial direction amounts to  
\begin{align}
i\rightarrow \infty, \ N\rightarrow \infty, \ \frac{i}{N}\equiv \frac{x}{L}={\cal O}(1) \  . 
\end{align}
In this situation, it is not hard to show that one can introduce 
\begin{align}
w=e^{-\frac{2\pi (r+ix)}{L}},   \   \bar w= e^{-\frac{2\pi (r-ix)}{L}} \ ,
\end{align}
and generalizes the Fredholm determinant formula to the following factorized form
\begin{align}
\langle \Omega_{NS}|\hat \sigma(r,x)\hat \sigma(0)|\Omega_{NS}\rangle=e^{-\frac{\pi r}{4L}}\det (1+{\cal  K}_{ij}(w))\det (1+{\cal  K}_{ij}(\bar w)) \ . \label{eq:Fredholmgeneral}
\end{align}
This is again consistent with the standard CFT formula~\cite{DiFrancesco:1987cx,DiFrancesco:1997nk}.

\section{Comparison with literature and further comments}\label{sec:compare}
Before finishing the work, let's compare with the literature and make the following comments. 

First, in the introduction, we mentioned that the massless limits of the finite volume massive form factors in~\cite{Fonseca:2001dc} were rarely mentioned, so the first thing we would like to comment are the leg functions in that reference. In the notation of~\cite{Fonseca:2001dc}, the function $\kappa(\theta)$ reads
\begin{align}
\kappa(\theta)=\int_{-\infty}^\infty\frac{d\theta'}{2\pi} \frac{1}{\cosh(\theta-\theta')}\ln \frac{1-e^{-\mu L \cosh \theta'}}{1+e^{-\mu L\cosh \theta'}} \ ,
\end{align}
where $\mu >0$ is the fermion mass.  At any fixed $\theta$, the $\mu \rightarrow 0$ limit diverges logarithmically. However, in order to reach the CFT limit, one must fix the quantization levels and the radius, then send $\mu$ to zero, which means $\theta \rightarrow \pm \infty $ and suppresses the $\theta'={\cal O}(1)$ region. On the other hand, when $\theta'-\theta={\cal O}(1)$, then the denominator is no longer large, and the $\mu$ in the exponential is also compensated by the $\mu L  \cosh \theta'\sim \mu L \cosh \theta $, which is $O(1)$ in the scaling limit. To magnify the $\theta'-\theta={\cal O}(1)$ region, we change the variable to the momentum space, and denote $p=\mu\sinh \theta, E=\mu\cosh \theta$
\begin{align}
&\kappa(EL,\mu L)=\int_{-\infty}^{\infty} \frac{dk}{2\pi \sqrt{k^2+\mu^2}} \frac{\mu^2}{E\sqrt{k^2+\mu^2}-pk}\ln \frac{1-e^{-L \sqrt{k^2+\mu^2}}}{1+e^{-L\sqrt{k^2+\mu^2}}}  \nonumber \\ 
&=\int_\mu^{\infty}\frac{dE'}{\pi\sqrt{E'^2-\mu^2}} \frac{EE'}{E'^2+E^2-\mu^2}\ln \frac{1-e^{-LE'}}{1+e^{-LE'}}  \ .
\end{align}
The first line is essentially the integral given in~\cite{Gabai:2019ryw}, while in the second line, we have averaged with the $k\rightarrow-k$ integrand and changed to $E'=\sqrt{k^2+\mu^2}$. In this form, it finally becomes transparent that there is a non-trivial $\mu\rightarrow 0^+$ limit at fixed $EL$
\begin{align}
\kappa(EL,0)=\int_0^{\infty}\frac{dE'}{\pi} \frac{E}{E^2+E'^2}\ln \frac{1-e^{-LE'}}{1+e^{-LE'}} = \frac{EL}{\pi}\int_0^{\infty} \frac{dt}{t^2+(EL)^2}\ln \frac{1-e^{-t}}{1+e^{-t}} \ . 
\end{align}
This is exactly the Binet's integral Eq.~(\ref{eq:binet}). As such, at the quantization levels $EL=2\pi |n| $ of the R sector, and at $EL=2\pi |m|$ ($m\in Z+\frac{1}{2}$) for the NS sector, the leg functions in~\cite{Fonseca:2001dc} evaluate to \begin{align}
&\frac{e^{\kappa(\theta_n)}}{\sqrt{\mu L\cosh \theta_n}} \rightarrow \frac{\Gamma(n+\frac{1}{2})}{\sqrt{2\pi} \Gamma(n+1)}=\tau(n) \ , \\ 
&\frac{e^{-\kappa(\theta_m)}}{\sqrt{\mu L\cosh \theta_m}} \rightarrow \frac{\sqrt{2\pi}\Gamma(m+1)}{2\pi m \Gamma(m+\frac{1}{2})}=\tau\left(m-\frac{1}{2}\right) \ ,
\end{align}
where $\tau$ is given by Eq.~(\ref{eq:deftau}). The above exactly reproduce the dressing functions in Eq.~(\ref{eq:formCFT1}) and Eq.~(\ref{eq:formCFT2}). For the zero mode, instead one needs 
\begin{align}
    \kappa(\mu L,\mu L)=\frac{1}{\pi}\int_1^{\infty} \frac{dt}{t\sqrt{t^2-1}}\ln  \frac{1-e^{-\mu Lt}}{1+e^{-\mu Lt}}=\frac{1}{2} \ln \mu L+{o}(1) \ , \ \frac{e^{\kappa(0)}}{\sqrt{\mu L}} \rightarrow 1  \ , 
\end{align}
again consistent with the $d(1)=1$. Two particle functions clearly simplify to the $\Pi$ factors in Eq.~(\ref{eq:formCFT1}) and Eq.~(\ref{eq:formCFT2}). Moreover, in Appendix.~\ref{sec:expandgI}, we show that normalization function $\bar \sigma S(\mu L)$ of~\cite{Fonseca:2001dc} has the following small $\mu$ expansion
\begin{align}
\bar \sigma S(\mu L)=\left(\frac{2\pi}{L}\right)^{\frac{1}{8}}\exp \bigg(\frac{\ln 2}{2\pi }\mu L-\frac{\ln^2 2}{4\pi^2}(\mu L)^2-\frac{\zeta_3}{16\pi^3}(\mu L)^3+{\cal O}((\mu L)^4)\bigg) \ . 
\end{align}
Given this, we conclude that the massless limit of the massive matrix elements in~\cite{Fonseca:2001dc}, up to overall phase factors, exactly reduce to $\left(\frac{2\pi}{L}\right)^{\frac{1}{8}}$ multiplying our Eq.~(\ref{eq:formCFT1}) and Eq.~(\ref{eq:formCFT2}) for $\hat \sigma$. The normalization of~\cite{Fonseca:2001dc}  ensures the small-$r$ limit of the massless two point correlator to be $r^{-\frac{1}{4}}$ without extra factors. 

Second, in the introduction, we have mentioned that in~\cite{Gehlen_2008}, the authors have omitted the $g=1$ point in their discussions, but this does not rule out the possibility that the results therein also apply at the $g=1$ point by taking limit. Let's show that this is indeed the case.  The major quantity to check are the products of their leg function. In our notation, the leg function in~\cite{Gehlen_2008} for periodic momenta reads 
\begin{align}
\frac{1}{N\epsilon(\beta)}e^{-\eta(\beta)}=\frac{1}{N\epsilon(\beta)}\frac{\prod_{\beta'\in S_+}(\epsilon(\beta)+\epsilon(\beta'))}{\prod_{z' \in S_-}\left(\epsilon(\beta)+\epsilon(z')\right)} \  ,  \  \beta \ne 1 \ .
\end{align}
For $\beta=1$, if one start from the $g>1$ region and then take the $g\rightarrow1^+$ limit, then the result is finite
\begin{align}
\frac{1}{N\epsilon(1)}e^{-\eta(1)}=\frac{2}{N}\frac{\prod_{\beta \ne 1} \epsilon(\beta)}{\prod_{z\in S_-}\epsilon(z)} \ . 
\end{align}
Let's see if they reproduces our $|d(\beta)|^2$ and $|d(1)|^2=1$. We first check $\beta=1$. One has 
\begin{align}
\frac{2}{N}\frac{\prod_{\beta \ne 1} \epsilon(\beta)}{\prod_{z\in S_-}\epsilon(z)}=\frac{2}{N}\times(-i)\frac{\prod_{\beta \ne 1}(1-\beta)}{\prod_{z\in S_-}(1-z)}\frac{\prod_{z\in S_-} \sqrt{z}}{\prod_{\beta \in S_+}\sqrt{\beta}}=\frac{2}{N}(-i)\frac{N}{2}\frac{(-1)^{\frac{N}{2}}}{i(-1)^{\frac{N-2}{2}}}=1 \ . 
\end{align}
Here, we have used the unrestricted products when showing the Eq.~(\ref{eq:refg}), and the products in Eq.~(\ref{eq:productsimple}). As expected, the $|d(1)|^2=1$ is reproduced. For $\beta \ne1$, notice 
\begin{align}
\sqrt{\beta}+\sqrt{\beta'}-\frac{1}{\sqrt{\beta}}-\frac{1}{\sqrt{\beta'}}=\left(\sqrt{\beta}+\sqrt{\beta'}\right)\left(1-\frac{1}{\sqrt{\beta}\sqrt{\beta'}}\right) \ , 
\end{align}
and using the fact that $\sqrt{\frac{1}{\beta}}=-\frac{1}{\sqrt{\beta}}$ in the $(0,2\pi)$ branch when $\beta \ne1$, one has 
\begin{align}
&\frac{\prod_{\beta'\in S_+}(\epsilon(\beta)+\epsilon(\beta'))}{\prod_{z' \in S_-}\left(\epsilon(\beta)+\epsilon(z')\right)}=\frac{\sqrt{\beta}-1}{\sqrt{\beta}+1}p(\beta)^2 \ , \\
&\frac{1}{N\epsilon(\beta)}e^{-\eta(\beta)} \rightarrow\frac{i \sqrt{\beta}p(\beta)^2}{N(1+\sqrt{\beta})^2}=\bigg|\frac{\sqrt{\beta}p(\beta)}{\sqrt{N}(1+\sqrt{\beta})}\bigg|^2=|d(\beta)|^2 \ , 
\end{align}
where we have used the definitions of $p(\beta)$ and $d(\beta)$ in Eq.~(\ref{eq:deffb}), Eq.~(\ref{eq:defdressb}), and the fact that $p(\beta)=e^{-\frac{i\pi}{4}}|p(\beta)|$. On the other hand, for $z\in S_-$, one can also compute 
\begin{align}
\frac{1}{N\epsilon(z)}e^{\eta(z)}=\frac{1}{N\epsilon(z)}\frac{\prod_{z' \in S_-}\left(\epsilon(z)+\epsilon(z')\right)}{\prod_{\beta'\in S_+}(\epsilon(z)+\epsilon(\beta'))}=\frac{\sqrt{z}}{N(\sqrt{z}-1)^2}\frac{1}{-ip(z)^2}=|d(z)|^2 \ , 
\end{align}
where we have again used $p(z)=e^{-\frac{i\pi}{4}}|p(z)|$,  Eq.~(\ref{eq:deffz}) and Eq.~(\ref{eq:defdressz}). As such, we have found that all the leg functions of the $g\rightarrow 1^+$ limit in~\cite{Gehlen_2008} agree with our results. For the two-particle functions, it is also easy to check that they agree with our expressions. Finally, for the overall normalization factor $\xi\xi_T$, notice that the $\xi=(g^2-1)^{\frac{1}{4}}$ again cancels with the $(2\epsilon(1))^{\frac{1}{4}}$ in the denominator, and the $g\rightarrow1^+$ limit is again finite. Moreover, using the relations above, one can compute
\begin{align}
&\lim_{g\rightarrow 1^+}\xi \xi_T=\bigg(-\frac{\prod_{z\in S_-}(1-\sqrt{z})^2}{\prod_{\beta \ne 1}(1-\sqrt{\beta)^2}}p^2(1)\prod\frac{p^2(z)}{p^2(\beta)}\bigg)^{\frac{1}{4}}\nonumber \\ 
&=\left(\frac{2}{N}\right)^{\frac{1}{4}}\exp \bigg(\frac{1}{2}\sum_{z\in S_-}\hat \Gamma_N(z)-\frac{1}{2}\sum_{\beta \in S_+, \beta \ne 1}\hat \Gamma_N(\beta)\bigg)=|\langle \Omega_R|\Omega_{NS}\rangle|^2 \ , 
\end{align}
where we have used the relation $p(1)q(1)=-2i$ in Eq.~(\ref{eq:pq1}). As such, all the form factors squared in~\cite{Gehlen_2008,Iorgov:2010yv}, in the proper $g\rightarrow 1^+$ limit approached from the disordered side, agree with the results derived directly at the critical point.

After comparing with the literature, we would like to make a few more comments on possible future directions and generalizations.
\begin{enumerate}
\item First, we should emphasize that although we believe that the Fredholm determinant Eq.~(\ref{eq:Fredholmscale}) leads exactly to the CFT correlator through the formula
\begin{align}
\det (1+{\cal K})=(1-w)^{-\frac{1}{8}} \ , \label{eq:identityfred}
\end{align}
we have not yet verified this analytically. However, we have noticed that the expansion of this determinant is similar to the infinite-volume form factor expansion~\cite{Lyberg_2007} of the Ising chain or diagonal Ising correlators away from the critical point, and at zero or very small distances. For example,  the first two terms in the expansion equal to
\begin{align}
1+{\rm Tr} {\cal K}=1+\sum_{i=1}^\infty \frac{\Gamma^2(i+\frac{1}{2})}{2\pi\Gamma^2(i+1)}w^i=\frac{K(w)}{\pi }+\frac{1}{2} \ , 
\end{align}
where $K(w)$ is the complete Elliptic integral of the second Kind. The $\frac{K(w)}{\pi}$ is exactly half of the one particle form factor contribution $f^{(1)}_{0,0}(t)$ to the diagonal correlator at zero separation~\cite{Boukraa:2006bt}. If such relations could be established at high orders, then the identity Eq.~(\ref{eq:identityfred}) becomes understandable: $\sigma_{00}^2=1$, while the form factor expansion in~\cite{Boukraa:2006bt} contains an overall $(1-t)^{\frac{1}{4}}$. This implies that all the form factors at zero separation must sum to $(1-t)^{-\frac{1}{4}}$. It is also possible that differential equations in $w$ could be established, similar to the Painleve VI for the parameter derivatives~\cite{10.3792/pjaa.56.405}.

    \item Second, although in this work we only checked the two-point correlator, given the finite-volume form-factors, it should be possible to show the convergence of the scaling limit and establish determinantal representations for multi-point spin-correlators at non-vanishing Euclidean time separations,  in a way similar to the massive version in the infinite volume~\cite{McCoy:1977er,PALMER1981329}. What remains not clear at the moment, is how to use the form factor representations, to show the resulting correlators indeed agree with the known CFT results. 
    
    On the other hand, it was proven recently that the spin correlators of the 2D critical {\it Ising model} on a {\it torus} indeed converge to the CFT version~\cite{bayraktaroglu2025criticalisingcorrelationstorus}, using more subtle methods that are not based on form factors. Meanwhile, finite-volume form factors of the critical Ising model can still be established using the strategy of~\cite{cmp/1103901557,cmp/1103904079} or in the formalism of~\cite{Iorgov:2010ie} based on elliptic Cauchy determinants, by noticing similar near Cauchy property at the critical point\footnote{The author thanks Oleg Lisovyi for pointing this out in private communications.}. They are expected to converge to the same set of CFT matrix elements, leading to identical form factor representations of rescaled correlators as obtained from the Ising chain. As such, by comparing the form factor representations and the known CFT results allows to establish many Fredholm determinant identities that are not obvious at the first glance at all. Clearly, it is also interesting to see how the form factors can be summed to the more general toroidal correlators at a finite temperature. 
    \item Third, we comment that the finite volume form-factors are also known explicitly for more general $XY$ spin-chains in a transverse field~\cite{Iorgov_2011} away from the critical points. Either by taking the appropriate $XX$ limit ($\kappa=h=0$) or by deriving the finite-volume form factors directly at that point which is technically possible, one should be able to obtain form factors of certain vertex operators in the $c=1,\ \beta^2=4\pi$ Sine-Gordon CFT in the massless Dirac fermion basis.  
    \item  Finally, we should mention that for integrable lattice models solved by the Algebraic Bethe-Ansartz (ABA) such as the $XXZ$ spin-chains, there are also discussions on the large distance properties  based on form-factors~\cite{Dugave:2014xka,Gohmann:2018cfz,Gohmann:2019vgl,Kozlowski:2019bng,Gohmann:2023lbr}. These discussions mostly concern the infinite-volume limit at zero or non-zero temperatures, but the region $-1\le \Delta<1$ with CFT-like asymptotics are covered.

It is interesting to see how to extract from the $XXZ$ spin-chain, the CFT matrix elements in more general $c=1$ Sine-Gordon CFTs in the ABA-type massless basis, by starting with a finite volume, then taking the finite-volume scaling limit. It could be expected that going to a finite volume would simplify a bit the analysis aspects (taking limits) of the massless form factors. Such matrix elements might also be related to the finite temperature results in~\cite{Babenko:2020spo,Gohmann:2023lbr,Gohmann:2023fcl}. Showing the convergence to the cylindrical CFT correlators is expected to be more challenging than to obtain the form factors.
\end{enumerate}

\acknowledgments
The author thanks Zoltán Bajnok for constructive discussions and introduction to the references~\cite{Yurov:1991my,Fonseca:2001dc}. In particular, the author was alerted that there seems to be no explicit formulas for the CFT matrix elements in~\cite{Yurov:1991my} after discussing with him. 

The results Eq.~(\ref{eq:formCFT1}), Eq.~(\ref{eq:formCFT2}) and Eq.~(\ref{eq:phases})  for the CFT matrix-elements were double checked by two less rigorous semi-infinite versions of this work, corresponding to taking the scaling limit at two different early stages. Notes for these computations can be provided upon request. The product formulas were checked by Zoltán Bajnok against the algorithm of~\cite{Yurov:1991my} with his private codes up to phase conventions.

\appendix

\section{The $s\rightarrow0^+$ expansion of a crossover function} \label{sec:expandgI}

In this appendix, we compute analytically the small $s\rightarrow0^+$ expansion of the function $g_I(s)$ of~\cite{Sachdev:1995bf} as an application of the Barnes integral Eq.~(\ref{eq:Barnessymetric}). We start from the definition with $s>0$
\begin{align}
&-f_I(s)=\frac{1}{\pi}\int_0^{\infty}dy \ln \frac{1-e^{-\sqrt{y^2+s^2}}}{1+e^{-\sqrt{y^2+s^2}}} \ , \\
&\ln g_I(s)=\int_s^1 \frac{ds'}{s'} \left((f'_I(s'))^2-\frac{1}{4}\right)+\int_1^{\infty}\frac{ds'}{s'} (f_I'(s))^2 \ . 
\end{align}
To perform the integral, it is convenient to use the following Mellin-Barnes representation of $f_I'(s)$
\begin{align}
&-f'_I(s)=\int_{{\rm Re}(u)>0}\frac{du}{2\pi i }s^{-u}{\cal M}(u) \ , \\
&{\cal M}(u)=\frac{\left(2^{u+1}-1\right) \Gamma^2 \left(\frac{u+1}{2}\right)\zeta(u+1)}{2 \pi } \ . 
\end{align}
The above can be established by expanding the logarithm into the Taylor series
\begin{align}
\ln \frac{1-e^{-\sqrt{y^2+s^2}}}{1+e^{-\sqrt{y^2+s^2}}}=-2\sum_{k=0}^{\infty}\frac{e^{-(2k+1)\sqrt{y^2+s^2}}}{2k+1} \ , 
\end{align}
and Mellin-transform the individual terms. It is easy to see that 
\begin{align}
{\rm Res} \bigg({\cal M}(u)\bigg)(u=0)=\frac{1}{2} \ , 
\end{align}
consistent with the subtraction term $-\frac{1}{4}$. Given the above, it is then straightforward to establish the following double Barnes integral
\begin{align}
&\ln g_I(s)=\int \int_{{{\rm Re}(u)-1}<{\rm Re}(z)<0<{\rm Re}(u)<1}\frac{dz du }{(2\pi i)^2}\frac{s^{-z}-1}{z}{\cal M}(u){\cal M}(z-u) \nonumber \\ 
&+\frac{1}{2}\int_{-1<\rm Re(u)<0} \frac{du}{2\pi i }\frac{s^{-u}-1}{u}{\cal M}(u)+\int \int_{0<{\rm Re}(u)<{\rm Re}(z)<1}\frac{dz du }{(2\pi i)^2}\frac{{\cal M}(u){\cal M}(z-u)}{z} \ .  
\end{align}
Here, the first two terms are due to the integral in the $(0,1)$ region, while the last term is due to the integral in the $(1,\infty)$ region. The $-\frac{1}{4}$ subtraction is achieved by appropriate contour shifting. We now shift the $z$ contour in the last term to the ${\rm Re}(z)<0$ region to cancel the $s$ independent terms. There are $z=0$ and $z=u$ two poles. Using the fact that 
\begin{align}
{\rm Res} \left(\frac{{\cal M}(u)}{u}\right)(u=0)=0 \ , 
\end{align}
one finally obtains 
\begin{align}
&\ln g_I(s)=\int \int_{{{\rm Re}(u)-1}<{\rm Re}(z)<0<{\rm Re}(u)<1}\frac{dz du }{(2\pi i)^2}\frac{s^{-z}}{z}{\cal M}(u){\cal M}(z-u) \nonumber \\ 
&+\frac{1}{2}\int_{-1<\rm Re(u)<0} \frac{du}{2\pi i }\frac{s^{-u}}{u}{\cal M}(u)+\int_{0<{\rm Re}(u)<1} \frac{du}{2\pi i }{\cal M}(u){\cal M}(-u)\ . \label{eq:gIbarnes}
\end{align}
Now, using the functional equation of the $\zeta$ function, it is straightforward to show that 
\begin{align}
{\cal M}(u){\cal M}(-u)\equiv  \left(2-2^{u}\right) \left(2-2^{-u}\right) \Gamma (u) \Gamma (-u)\zeta (u) \zeta (-u) \ .
\end{align}
As such, the last term of Eq.~(\ref{eq:gIbarnes}) exactly leads to the Barnes-$G$ product using the Barnes integral Eq.~(\ref{eq:Barnessymetric}). On the other hand, in the first two terms, one can shift the contours to the left to perform systematically the small $s$ expansion. The results of the expansion up to the third order read
\begin{align}
\ln g_I(s)\bigg|_{s\rightarrow 0^+}=\ln \bigg(\pi^{\frac{1}{4}}G\left(\frac{1}{2}\right)G\left(\frac{3}{2}\right)\bigg)+\frac{\ln 2 }{\pi}s-\frac{\ln^22}{2\pi^2}s^2-\frac{\zeta_3}{8\pi^3}s^3+{\cal O}(s^4) \ . \label{eq:expandgI}
\end{align}
In particular, the leading term exactly reproduces the constant ${\cal C}_1$ extracted numerically in~\cite{Sachdev:1995bf}. It is also in agreement with the Eq.~(5.19) of~\cite{Bugrij:2001nf}. At higher orders, there will be $\zeta_{2k+1}\zeta_{2l+1}$ type terms. The expansion has a convergence radius equals to $\pi$.

Given the expansion of $\ln g_I(s)$, the expansion of the function $S(\mu L)$ in~\cite{Fonseca:2001dc} can also be found using the relation
\begin{align}
\ln S(\mu L)=\frac{1}{2}\ln g_I\left( \mu L\right)-\frac{1}{8}\ln \mu L \ , 
\end{align}
which is equivalent to Eq.~(A13) of~\cite{Fonseca:2001dc}. Especially, using the expansion Eq.~(\ref{eq:expandgI}), in the massless limit the $\bar \sigma \mu ^{\frac{1}{8}}S(\mu L)$ is finite 
\begin{align}
\lim_{\mu \rightarrow 0}\mu ^{\frac{1}{8}}2^{\frac{1}{12}}e^{-\frac{1}{8}}A^{\frac{3}{2}} S(\mu L)=\frac{2^{\frac{1}{12}}e^{-\frac{1}{8}}A^{\frac{3}{2}}}{L^{\frac{1}{8}}}(\pi e)^{\frac{1}{8}}2^{\frac{1}{24}} A^{-\frac{3}{2}}=\left(\frac{2\pi}{L}\right)^{\frac{1}{8}} \ .
\end{align}
Using the Eq.~(2.15) of~\cite{Fonseca:2001dc} and the massless limits of their leg functions discussed in Sec.~\ref{sec:compare}, this implies that the overall normalization of the spin operator in~\cite{Fonseca:2001dc} is exactly $\left(\frac{2\pi}{L}\right)^{\frac{1}{8}}$ times our $\hat \sigma$. In particular, at small $r$, the two point correlator approaches a simple power-law $r^{-\frac{1}{4}}$ without extra factors.

\section{The integral Eq.~(\ref{eq:Claiinte1})} \label{sec:integral}
In this appendix, we prove the integral formula Eq.~(\ref{eq:Claiinte1}). We first change the variable from $\sqrt{t} \rightarrow t$, and introduce another integral representation for $(1-t)(\ln t)^{-1}$, leading to
\begin{align}
&I_0=\frac{1}{2}\int_0^1\frac{dt}{\ln t}\frac{1-t}{(1+t)^2} =-\frac{1}{2}\int_0^1dx\int_0^1 dt \frac{t^x}{(1+t)^2} \ .
\end{align}
It convergences absolutely. The $t$ integral can be integrated to a special case of the Gauss Hypergeometric function and can be expressed in terms of the digamma function, leading to
\begin{align}
&\int_0^1 dt  \frac{t^x}{(1+t)^2}=\frac{1}{2}+\frac{x}{2}\left(\psi\left(\frac{1+x}{2}\right)-\psi\left(1+\frac{x}{2}\right)\right) \ ,  \\
&I_0=-\frac{1}{4}-\frac{1}{4}\int_0^1 dx x\left(\psi\left(\frac{1+x}{2}\right)-\psi\left(1+\frac{x}{2}\right)\right) \ . 
\end{align}
The integral here is similar to that of the Ising connecting computation~\cite{Tracy:1990tn}. Using $\psi(x)=(\ln \Gamma(x))'$ and partial integrating as in~\cite{Tracy:1990tn}, one has 
\begin{align}
I_0=\frac{1}{2}\ln \frac{\Gamma \left(\frac{1}{2}\right)}{\Gamma \left(1\right)}+\ln \frac{\Gamma_2^2(1)}{\Gamma_2\left(\frac{1}{2}\right)\Gamma_2\left(\frac{3}{2}\right)}=\ln \bigg(\pi^{\frac{1}{4}} G\left(\frac{1}{2}\right) G\left(\frac{3}{2}\right)\bigg)  \ .
\end{align}
Here we have used the relation $G({\cal Z})=\frac{1}{\Gamma_2({\cal Z})}$ between the Barnes $G$ function and the double gamma function $\Gamma_2({\cal Z})$. This is exactly Eq.~(\ref{eq:Claiinte1}).

\bibliographystyle{apsrev4-1} 
\bibliography{bibliography}

\end{document}